\documentclass[prl,reprint]{revtex4-1}

\usepackage{amsmath}
\usepackage{txfonts}
\usepackage{hyperref}
\usepackage{graphicx}
\usepackage{float}
\usepackage[caption=false]{subfig}
\usepackage{color}
\usepackage[normalem]{ulem}
\usepackage{bm}
\usepackage{todonotes}

\begin{document}
\title{Emergence of L\'evy walks from second order stochastic optimization}
\author{\L{}ukasz Ku\'smierz}
\affiliation{RIKEN Brain Science Institute, 2-1 Hirosawa, Wako, Saitama 351-0198, Japan}
\author{Taro Toyoizumi}
\affiliation{RIKEN Brain Science Institute, 2-1 Hirosawa, Wako, Saitama 351-0198, Japan}
\begin{abstract}
In natural foraging, many organisms seem to perform two different types of motile search:
directed search (taxis) and random search.
The former is observed when the environment provides cues
to guide motion towards a target.
The latter involves no apparent memory or information processing and
can be mathematically modeled by random walks.
We show that both types of search can be generated by a common mechanism 
in which L\'evy flights or L\'evy walks 
emerge from a second-order gradient-based search with noisy observations. 
No explicit switching mechanism is required -- 
instead, continuous transitions between the directed 
and random motions emerge depending on the Hessian matrix of the cost function. 
For a wide range of scenarios the L\'evy tail index is $\alpha=1$, 
consistent with previous observations in foraging organisms. 
These results suggest that adopting a second-order optimization method 
can be a useful strategy to combine efficient features of directed and random search.
\end{abstract}

\maketitle
Many organisms must actively search for resources 
in order to survive and produce offspring.
Foraging theory examines the various search strategies implemented by organisms
depending on their abilities and the environments in which they live.
In directed search, greater involvement of sensory and information 
processing abilities enable more complicated strategies.
In contrast, in the boundary case of a memoryless and senseless forager, 
the only option is to wander randomly in the environment (random search).
Even in this case, however, different strategies exist, 
depending on the character of the random motion.
A natural candidate model for the random strategy is Brownian motion 
that describes a wide range of natural phenomena, 
including the movement of inanimate particles under thermal noise. 
A prominent feature of Brownian motion is the linear growth of 
the variance of the position with time. 
However, empirical data indicate that for organisms the observed growth is often faster. 
L\'evy walks (LWs) 
\cite{shlesinger1982random,shlesinger1986levy,shlesinger1987levy} 
and similar L\'evy flights (LFs) 
\cite{mandelbrot1983fractal,shlesinger1986levy,metzler2000random} 
have been successfully applied to fit experimental data  
obtained from the movement patterns of many organisms and their cells, including 
T cells \cite{harris2012generalized}, 
microglia \cite{grinberg2011spreading}, 
starved slime mould (\textit{Dictyostelium discoideum})
\cite{takagi2008functional,reynolds2010can}, 
swarming bacteria \cite{ariel2015swarming}, 
fruit flies \cite{reynolds2007free}, 
honey bees 
\cite{reynolds2007displaced,wolf2016optimal},
wandering albatrosses
\cite{viswanathan1996levy,humphries2013new},
marine predators \cite{sims2008scaling}, 
and humans 
\cite{brockmann2006scaling,gonzalez2008understanding,rhee2011levy} 
(also in human's gaze \cite{brockmann2000ecology} and 
word association \cite{costa2009scale} trajectories). 
In many different random search scenarios, 
LWs and LFs have been shown to be advantageous over normal diffusion
\cite{viswanathan1999optimizing,buldyrev2001average,bartumeus2002optimizing,raposo2003dynamical,bartumeus2005animal,pasternak2009levy,raposo2011landscape,palyulin2014levy,humphries2014optimal,kusmierz2014first,kusmierz2015optimal}
and alternative superdiffusive models \cite{reynolds2009scale}.
These observations have led to the so-called L\'evy flight optimal foraging hypothesis,
which states that LFs (or LWs) represent evolutionary adaptations due to their
distinct advantages over other random search strategies
\cite{viswanathan1999optimizing,bartumeus2007levy,viswanathan2008levy}.

Recently this view has been disputed because none of the mentioned organisms is 
senseless and all of them are able to perform some forms of directed search (taxis), 
for example T cells and isolated bacteria perform chemotaxis
\cite{taub1993preferential,gunn1998chemokine,adler1966chemotaxis,berg1974chemotaxis,szurmant2004diversity},
whereas fruit flies perform phototaxis \cite{zhu2009peripheral},
geotaxis \cite{miller1999measuring}, and chemotaxis
\cite{duistermars2009flies,wasserman2013drosophila}.
Indeed, a number of studies have shown that
characteristics of LFs and LWs may emerge naturally on large scales 
from more realistic case specific models of movement
\cite{reynolds2015liberating}, 
including simple deterministic and semi-deterministic walks in complex environments 
\cite{boyer2004modeling,santos2007origin,viswanathan2008levy,reynolds2008deterministic,reynolds2012fitness,reynolds2014levy}, 
self-avoiding random walks 
\cite{shlesinger1983weierstrassian,reynolds2014mussels,sims2014hierarchical}, 
diffusion with a time-varying diffusion constant \cite{ott1990anomalous,srivastava2009temporal,salvador2014mechanistic}, 
and a multiplicative, self-accelerating process \cite{biro2005power,lubashevsky2009realization,reynolds2010can}.
It has also been suggested that in some species a power-law distribution 
of lengths of straight line segments of their movement patterns, 
a hallmark of LWs and LFs, 
is a consequence of either the Weber-Fechner law in odometry \cite{reynolds2013levy}, 
a power-law distribution of switching times between competing activities 
\cite{korobkova2004molecular,tu2005white,barabasi2005origin,matthaus2009coli,matthaus2011origin,reynolds2011origin}, 
or a so-called aerial lottery \cite{katul2005mechanistic,shaw2006assembling,reynolds2013beating}. 
Moreover, in some cases the apparent superdiffusive character of 
the population dynamics may be 
an artifact of averaging over an ensemble of the diffusive motions of individuals 
with diverse characteristics \cite{petrovskii2011variation}.


These studies suggest LWs and LFs naturally arise in many realistic biological settings 
but they do not argue why an apparent common behavior is observed across species and environments. 
Recently, a generalization of the LF optimal foraging hypothesis was proposed 
that explicitly combines directed and random search strategies. 
Specifically, an ad-hoc combination of taxis for choosing a direction and 
random, heavy-tailed distributed step-lengths was shown to be efficient 
under some search conditions \cite{pasternak2009levy,nurzaman2009yuragi}. 
In contrast, here we propose a novel mechanism by which 
LWs and LFs can emerge from a generic, locally optimal, directed search strategy. 
In our model the directed search is realized as a taxis driven by 
local observations of a cost function 
(e.g. a repellent concentration minus an attractant concentration) 
whose minima correspond to targets. 
Inspired by the second-order gradient-based optimization techniques known from computer science 
we assume that the search is based on noisy gradient and Hessian estimates. 
As we show below, this generically leads to heavy tails of the steps distribution. 
In contrast to previous models, our model predicts continuous crossover 
between random L\'evy searches and directed, deterministic taxis 
depending on the amount of information on the target location 
provided by observations.

This letter is organized as follows. 
First, we fix the notation and introduce a one-dimensional version of our model.
Next, we list different scenarios in which we are able to prove 
the existence of the heavy tails. 
We then discuss how the tails are affected by the landscape and 
the observation methods. 
Finally, we discuss a multidimensional generalization of 
the model followed by concluding remarks.

Let $(x_n)_{n=0}^{\infty}$ be a sequence generated by the Newton optimization rule
\begin{equation}
x_{n+1} = x_n + \Delta_n,
\label{eq:random-walk}
\end{equation}
with
\begin{equation}
\Delta_n = - \frac{f'(x_n) + \xi_G^{(n)} }{f''(x_n) + \beta + \xi_H^{(n)}},
\label{eq:def-delta}
\end{equation}
where the cost function $f{:}\mbox{ } \mathbb{R}\to\mathbb{R}$ is to be minimized. 
The rule with $\beta=\xi_G^{(n)}=\xi_H^{(n)}=0$ performs 
a gradient descent or a gradient ascent on $f$, depending on its curvature. 
A positive constant $\beta$ (damping) is added to the denominator 
in order to turn this algorithm into a minimizer. 
Note that the steepest descent method:
\begin{equation}
\tilde{\Delta}_n = - \frac{f'(x_n) + \xi_G^{(n)} }{\beta},
\label{eq:def-delta-1st}
\end{equation} 
is recovered from (\ref{eq:def-delta}) in the limit of $\beta\to\infty$ 
if $f$ has a bounded second derivative. 
Terms $\xi_G^{(n)}$ and $\xi_H^{(n)}$ account for noise: 
if the optimization is to be performed in the physical world, 
derivatives of $f$ are based on noisy measurements. 
Similarly, in many optimization problems solved on a computer, 
especially in machine learning, 
a function to be optimized is estimated with finite precision. 
With these definitions the sequence $(x_n)_{n=0}^{\infty}$ 
denotes a one-dimensional discrete-time random walk. 

Trajectories of LWs consist of linear segments 
(or instantaneous jumps in the case of LFs) $\Delta_n$, 
which are \textit{i.i.d.} random variables
(hence we omit the time-index $n$ in the discussion of distributions and write simply $\Delta$).
The probability density function (PDF) of $\Delta$
is characterized by heavy tails
i.e. for large $|z|$
\begin{equation}
\rho_{\Delta} (z) \sim |z|^{-1-\alpha},
\label{eq:heavy-tails-alpha}
\end{equation}
where the tail index $0<\alpha<2$. 
In the following we show that for a wide range of scenarios 
the random walk defined by (\ref{eq:random-walk}) and (\ref{eq:def-delta}) 
is equivalent to a (possibly inhomogeneous) LW or LF 
(depending on how it is mapped into a continuous time process \cite{zaburdaev2015levy}) 
with $\alpha=1$. 
We shall first analyze the case when $\beta=0$ 
and $f(x)=\mbox{const.}$, so that only noise is sampled. 
Assuming that the noise is Gaussian and that both $f'$ and $f''$ 
are measured independently and without bias, we can write 
\begin{equation}
\Delta = - \frac{\xi_G}{\xi_H}, 
\label{eq:quotient-gaussians}
\end{equation}
where $\xi_G$ and $\xi_H$ are independent Gaussian variables
with zero mean and standard deviations $\sigma_G$ and $\sigma_H$.
The reader can easily verify that $\Delta$ is in this case 
characterized by the Cauchy distribution
\begin{equation}
\rho_{\Delta}(z) = \frac{1}{\pi} \frac{\gamma}{\gamma^2 + z^2},
\label{eq:cauchy-dist}
\end{equation}
where $\gamma = \frac{\sigma_G}{\sigma_H}$.
Comparing (\ref{eq:cauchy-dist}) with (\ref{eq:heavy-tails-alpha}) 
we see that in our case $\alpha=1$.
More generally, let us assume that the numerator $X_G$ and denominator $X_H$ 
in (\ref{eq:def-delta}) are independent random variables 
with PDFs $\rho_{X_G}$ and $\rho_{X_H}$, respectively. 
This is the case if $\xi_G$ and $\xi_H$ are conditionally independent given 
the current position of the walker. 
The asymptotic form of the PDF of $\Delta = -X_G/X_H$ is given by
\begin{equation}
\begin{split}
\rho_{\Delta} (z) 
&= 
\int\limits_{-\infty}^{\infty} \mathrm{d}z_1 
\rho_{X_G}(z_1)
\int\limits_{-\infty}^{\infty} \mathrm{d}z_2 
\rho_{X_H}(z_2)
\delta\left(z + \frac{z_1}{z_2}\right) = 
\\ &=
\frac{1}{z^2} 
\int\limits_{-\infty}^{\infty} \mathrm{d}z_1
\rho_{X_G}(z_1) \rho_{X_H}\left( -\frac{z_1}{z} \right) |z_1|
= 
\frac{\langle |X_G| \rangle \rho_{X_H}(0)}{z^2} + o\left({z^{-2}}\right)
,
\end{split}
\label{eq:proof-1d-general}
\end{equation}
where the last equality holds if 
$\langle |X_G| \rangle \equiv \int \rho_{X_G}(z) |z| \mbox{d}z < \infty $
and
$0 < \rho_{X_H}(0) = \lim_{z \to 0^{\pm}} \rho_{X_H}(z)< \infty$.
The condition $\langle |X_G| < \infty$ is equivalent to the statement that
the tails of $\rho_{X_G}(z)$ decay faster than $z^{-2}$. 
If this condition is not fulfilled the appearance of heavy tails in the distribution of $\Delta$ is trivial. 
In our case, however, the heavy tails of $\rho_{\Delta}$ appear due to a non-zero 
probability of $X_H$ being arbitrarily close to zero. 
The described mechanism is very general as it does not assume 
that the noise distribution has heavy tails. 
Intuitively, the division in (\ref{eq:def-delta}) takes the role of a noise amplifier.
Clearly, first order methods, such as the steepest descent (\ref{eq:def-delta-1st}), 
do not involve a division by a random variable and 
therefore do not generically lead to heavy tails. 

In the case of correlated $X_G$ and $X_H$ the presence of heavy tails cannot be ensured in general. 
For instance, if $X_G = - Y X_H$ for some random variable $Y$, then the resulting 
$\Delta$ has the same distribution as $Y$. 
However, as we will now show, the heavy tails are still present in the generic case 
of the bivariate normal distribution of $X_G$ and $X_H$:
\begin{equation}
\rho_{\bm{X}}(\bm{x}) = 
\frac{|\bm{P}|^{1/2}}{2\pi} 
\exp\left(
-\tfrac{1}{2} (\bm{x}-\bm{\mu} )^{\intercal}
\bm{P}
(\bm{x} - \bm{\mu})
\right),
\end{equation}
where $\bm{X} = \left( \begin{smallmatrix} X_G \\ X_H \end{smallmatrix} \right)$ 
is a two-dimensional Gaussian random vector, 
$\bm{\mu} = \left( \begin{smallmatrix} \mu_G \\ \mu_H \end{smallmatrix}\right)$ 
is a vector of its expected values, 
$\bm{P}=\left( \begin{smallmatrix}P_{11} & P_{12} \\ P_{21} & P_{22} \end{smallmatrix} \right)$ 
is a symmetric, positive-definite precision matrix, 
and $|\bm{P}|$ is its determinant. 
The PDF of $\Delta$ can be calculated as 
$
\rho_\Delta(z) = 
\int \mathrm{d}^2 \bm{x} \rho_{\bm{X}}(\bm{x}) \delta(z + \frac{x_1}{x_2}).
$
and in the special case of $\mu_G = \mu_H = 0$ 
simplifies to the shifted Cauchy distribution
\begin{equation} 
\label{eq:Gauss}
\rho_{\Delta} (z) = 
\frac{1}{\pi} \frac{|\bm{P}|^{1/2}}
{P_{22} - 2 P_{12} z + P_{11} z^2}.
\end{equation}
In general, $\rho_\Delta$ takes the form
\begin{equation} 
\rho_{\Delta} (z) = 
\frac{|\bm{P}|^{1/2}}{2\pi z^2} 
\int\limits^{\infty}_{-\infty} \mathrm{d}x 
|x|
\exp\left(
-\tfrac{1}{2} 
(\bm{\tilde{x}} - \bm{\mu} )
^{\intercal}
\bm{P} 
(\bm{\tilde{x}} - \bm{\mu} )
\right)
=
\frac{I(z)}{z^2}
,
\end{equation}
where 
$\bm{\tilde{x}} = \left(\begin{smallmatrix} x\\ -x/z \end{smallmatrix}\right)$. 
Since $0 < \lim_{z \to \infty} I(z) < \infty$, we see that yet again
$\rho_{\Delta}(z)\sim 1/z^2$ for large $z$. 

\begin{figure}[t]
\centerline{\includegraphics[width = 1.1\linewidth,trim = {0 0 1cm 0},clip]{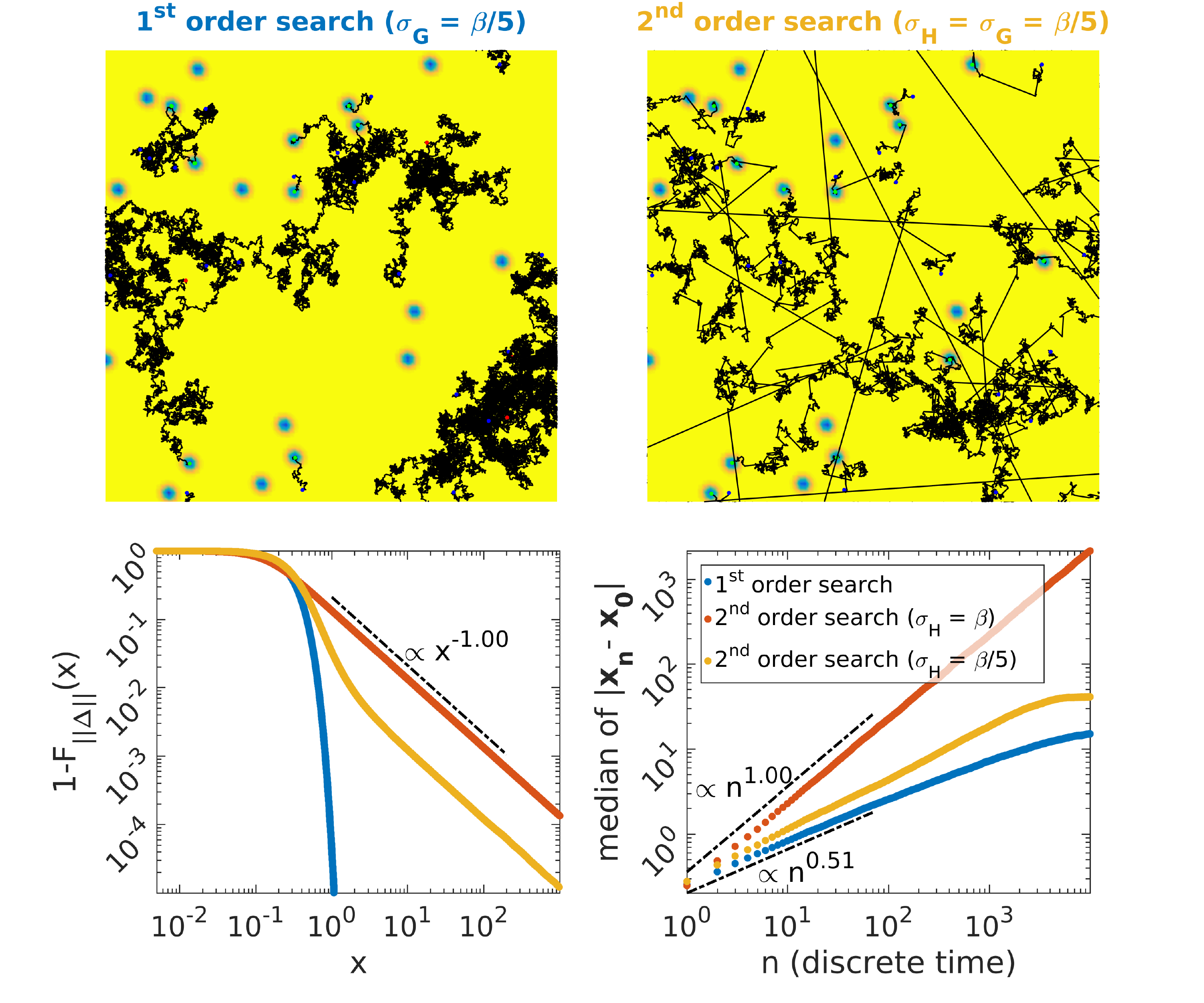}}
\caption{
An example of search processes in an unbounded 2-dimensional space.
The optimized function $f$ is a sum of $20$ Gaussians uniformly distributed 
within a $100\times100$ cell, which is periodically 
repeated across an infinite space.  
The targets are sparse so that in most places $f$ is flat and does not provide any
information about the position of the targets due to measurement noise.
The top two plots represent a cell of $f$ 
(yellow corresponds to high values, blue to low values) and $20$ exemplary trajectories of
searches starting from randomly chosen positions within the cell 
(blue dots) and finishing at a target (green dots)
or at some random position without finding the target (red dots) due to the time limitation ($10^4$ steps).
The bottom left plot depicts the distributions of jump lengths averaged over time and an ensemble of $10^4$ trajectories.
The second-order search produces a power-law tail with an exponent $\alpha \approx 1$. 
The bottom right plot shows the scaling of a displacement with time.
As expected, for short times the first order search leads to a diffusive behavior which scales as $n^{1/2}$,
whereas the second-order search with strong noise $\sigma_H$ leads to a superdiffusive behavior which scales as $n$.
For longer times the median displacement saturates due to trapping at the targets. %
}
\label{fig:random-gauss-2d}
\end{figure}

It may now seem like the second-order methods should always lead to LFs or LWs given noisy observations,
which might prevent them from being an efficient search strategy. 
However, this is not the case if the regularization factor $\beta$ and 
curvature $f''(x)$ in the denominator strongly temper heavy tails in (\ref{eq:def-delta}).
For example, in the case of independent Gaussian $\xi_G$ and $\xi_H$, the large $z$ limit of the 
cumulative distribution of the step size in (\ref{eq:proof-1d-general}) is given by 
\begin{equation}
P\left(|\Delta| > z \right)
\approx 
\frac{2\langle|X_G|\rangle \rho_{\xi_H}(-\beta-f''(x)) }{z}
=\frac{2\langle|X_G|\rangle}{\sqrt{2\pi}\sigma_H z} 
e^{-\frac{c^2}{2}} 
\label{eq:tempered-tails}
\end{equation}
with $c=\frac{\beta+f''(x)}{\sigma_H}$. 
Namely, heavy tails are still present,
but they are suppressed by the exponential factor $e^{-c^2/2}$.
Thus, for $|\beta+f''(x)|\gg\sigma_H$ the probability of large random 
displacements is extremely low. 
Equation (\ref{eq:tempered-tails}) provides a hint as to why the noisy 
second-order search may be efficient: 
if $\beta$ and $\sigma_H$ are chosen such that $c\approx0$ at the maxima,
where $f''$ is negative, 
and $c\gg 1$ at the minima, where $f''$ is positive, 
heavy tails are present in the vicinity of the maxima facilitating fast escapes, 
whereas around the minima heavy tails are strongly suppressed, 
allowing for an effective local exploration. 

We now address the question of how the method of estimating $f'$ and $f''$
from noisy measurements of $f$ can influence our results.
The simplest possible model in 1D consists of three observations.
Let us assume that the observations are performed at
$x_0 - \delta x$, $x_0$, and $x_0+\delta x$,
resulting in the following noisy measurements:
$y_- = f(x_0-\delta x) + \xi_-$,
$y_0 = f(x_0) + \xi_0$,
$y_+ = f(x_0+\delta x) + \xi_+$,
where $\xi_{\bullet}$ represent multivariate Gaussian noise.
If we assume that $\delta x$ is small enough we can write
the following formulas for the maximum likelihood estimates of the first two derivatives:
\begin{equation}
\begin{cases}
\hat{f}'(x_0) = \frac{y_+ - y_-}{2\delta x} \approx f'(x_0) + \xi_G,\\
\hat{f}''(x_0) = \frac{y_- + y_+-2 y_0}{\delta x^2} \approx f''(x_0) + \xi_H,\\
\end{cases}
\end{equation}
where $\xi_G = \frac{\xi_+ - \xi_-}{2\delta x}$ and
$\xi_H = \frac{\xi_+ -2 \xi_0 +\xi_-}{\delta x^2}$.
Hence, 
$\xi_G$ and $\xi_H$ are two jointly Gaussian random variables. 
As shown in (\ref{eq:Gauss}), this generally 
yields the LW or LF with $\alpha=1$.
This reasoning is still valid in scenarios with more measurements,
whenever the desired estimates are based on linear combinations of noisy observations.
Note that the more measurements are used in the estimators,
the better the Gaussian model of noise.


Finally, we turn our attention to the multidimensional case.  
For simplicity we assume that the search space is $\mathbb{R}^D$ 
with $D\in\mathbb{N}$. The jump vector, in analogy to (\ref{eq:def-delta}), 
takes the following form 
(as before, since we focus on a single step, we omit the step index)
\begin{equation}
\boldsymbol{\Delta} = 
-\boldsymbol{A}^{-1} \nabla \hat{f}(\boldsymbol{x})= 
- \left(\boldsymbol{H} f(\boldsymbol{x}) + 
\beta \boldsymbol{I} +\boldsymbol{\xi_H}\right)^{-1} 
\left( \nabla f(\boldsymbol{x})+\boldsymbol{\xi_G}\right), 
\label{eq:def-delta-D}
\end{equation}
where $\boldsymbol{H} f(\boldsymbol{x})$ denotes the Hessian of $f$, 
$\boldsymbol{\xi_G}$ is a noise vector, 
and $\boldsymbol{\xi_H}$ is a symmetric noise matrix.　
Under mild conditions, in the limit of $D\to\infty$ the noise 
eigenvalues $\lambda(\boldsymbol{\xi_H})$ 
follow the Wigner semicircle distribution 
\cite{wigner1958distribution,voiculescu1991limit,schenker2005semicircle,erdHos2011universality}. 
If the curvature and damping are much weaker than noise, 
they can only influence the distribution insignificantly, 
so that 
$0< \rho_{\lambda(\boldsymbol{A})}(0)<\infty$ still holds. 
Let $\boldsymbol{Q}$ be an orthogonal matrix diagonalizing 
$\boldsymbol{A}$. 
The $k$-th component of 
$\boldsymbol{Q}^T \boldsymbol{\Delta} = -\boldsymbol{Q}^T \boldsymbol{A}^{-1} \boldsymbol{Q} \boldsymbol{Q}^T \nabla \hat{f}(x)$ 
is proportional to $1/\lambda(\boldsymbol{A})_k$ 
and thus, according to (\ref{eq:proof-1d-general}),
its distribution has the heavy tail $1/z^2$. 
We can thus conclude \cite{samorodnitsky1994stable,nolan1999fitting}
that the distribution of $||\boldsymbol{\Delta}||$ also has the heavy 
tail $1/z^2$. 
In the continuous time limit this leads to a superdiffusive, 
multidimensional LW or LF 
\cite{samorodnitsky1994stable,teuerle2009random,teuerle2012multidimensional,szczepaniec2014stationary,szczepaniec2015escape,dybiec2015escape}. 
Note that the components of $\boldsymbol{\Delta}$ are not independent and 
the spectral measure \cite{samorodnitsky1994stable,szczepaniec2015escape} 
takes a nontrivial form, which will be the subject of future studies. 
In contrast, if the shift of eigenvalue distribution related to 
the curvature and damping factor is strong enough, 
the tails can be cut off completely, 
due to the bounded support of the Wigner semicircle distribution. 
In this case the continuous time limit process corresponds to diffusive search. 
Although for any $D<\infty$ the cut-off formally disappears, 
this shows that the heavy tails can be strongly tempered by the damping factor and the curvature. 
Importantly, the heavy tails may be tempered in the directions of large curvatures while 
being preserved in the other directions, thus providing a flexible combination of 
random and directed search mechanisms. 

We finally test our results using computer simulations in a simple example 
of a search in an unbounded two-dimensional space, 
see Fig. (\ref{fig:random-gauss-2d}). 
First, the first-order search in (\ref{eq:def-delta-1st}) does not produce 
power-law tails and its searching trajectory corresponds to that of normal 
diffusion except near a target. 
The median distance from the initial position scales as 
$\sqrt{n}$ with time step $n$.
Next, as predicted by our theory, the second-order search with a sufficiently large 
Hessian noise ($\sigma_H \approx f''+\beta$) produces heavy tails of 
the jump length distribution. 
In this case, the median distance from the initial position scales linearly with time, 
which is a characteristic of the LF process with $\alpha=1$. 
If the Hessian noise $\sigma_H$ is a few times smaller than $f''+\beta$, 
the behavior of the second-order method is in the middle between the above two extreme cases. 
In this case, the heavy tail of the step size distribution is present but somewhat tempered, 
in line with (\ref{eq:tempered-tails}). 
As a consequence, the growth of median distance from 
the initial position is initially ${\sim}\sqrt{n}$, 
similarly to normal diffusion, but slowly builds up in time as 
${\sim}n$ due to infrequent LF-like jumps. 
Note that, in the current simulation setup, the median distance saturates 
before the generalized central limit theorem predicts its linear growth 
because the trajectories are trapped by the targets sooner. 
This indicates that the second-order method with an appropriate noise level 
(or the damping constant) can find a target faster than the first-order 
search or the ad-hoc combination of directed search with power-law step sizes.
The detailed analysis of the optimal choice of the damping factor will be given elsewhere but, 
intuitively, it is beneficial to perform the first-order search in the direction of a convex 
surface and perform the LF search along the direction of a concave or plateau surface.

To sum up, we analyze a stochastic version of Newton's optimization method. 
We argue that noise in the estimates of the Hessian 
leads to a heavy-tailed distribution of jumps, an indicator of LWs or LFs. 
We present conditions in $D=1$ and $D\to\infty$ under which 
the appearance of the heavy tails is guaranteed and 
we corroborate these findings with computer simulations in 
the biologically relevant case of $D=2$. 

Our model explains how a seemingly common behavior 
(LW with the fixed tail index $\alpha=1$) can emerge from a generic and 
locally optimal search strategy in the presence of noise. 
This proposal is consistent with the role of evolution and 
adaptation under selection pressures in acquiring an advantageous 
search strategy. However, unlike some earlier proposals 
\cite{viswanathan1999optimizing,viswanathan2008levy,wosniack2017evolutionary}
that explore over the entire range of plastic $\alpha$, 
our model only gives two possible rigid values for $\alpha$: 
$\alpha=1$ generally and $\alpha=2$ in the limit of $(f''+\beta)/\sigma_H \to \infty$.
Moreover, our approach is distinct from the L\'evy flight foraging hypothesis 
\cite{viswanathan1999optimizing}, 
because the possible evolutionary optimization is not carried within the 
family of random search strategies (e.g., over a range of $\alpha$) 
but explicitly involves directed searches. 
Indeed, multiple specific mechanisms shaping directed search have been 
previously shown to produce LWs with an $\alpha=1$, 
e.g., movements in narrow, 
confined environments \cite{reynolds2015liberating},
in bulk-mediated effective surface diffusion 
\cite{bychuk1994anomalous,reynolds2015liberating}, 
and in patchy environments if foragers use information about patch quality 
\cite{reynolds2012fitness}.
Our proposal has two advantages over such findings: 
First, the second-order gradient-based optimization method is a well-established 
generic search strategy that works efficiently in many different environmental conditions. 
Second, by including taxis, our model suggests a specific continuous crossover 
between the random and directed search strategies as we further discuss below.  

Our results suggest that some organisms may perform taxis 
according to Newton's (or some other second-order) optimization method, 
which should be possible to verify experimentally. 
The resulting random walks are inhomogeneous and anisotropic, 
with less jerky motion along directions with larger curvatures 
or weaker measurement noise. 
These are distinct features of our model that can be taken advantage of 
in experiments aiming to assess whether foraging organisms 
employ second order derivatives. 
The candidate organisms that use taxis and, in some conditions, perform 
LWs with $\alpha\approx1$ include microglia \cite{grinberg2011spreading}, 
\textit{Dictyostelium discoideum} \cite{takagi2008functional,reynolds2010can},
and \textit{Drosophila} \cite{reynolds2007free}.
The strategy we introduce combines the characteristics of
two algorithms that are known to be efficient in directed 
(second-order optimization) and random (LFs or LWs) search scenarios. 
This method should therefore perform well in a broad range of 
scenarios of stochastic optimization, 
which may be of interest for the machine learning community. 


\bibliography{citations} 

\begin{thebibliography}{84}%
\makeatletter
\providecommand \@ifxundefined [1]{%
 \@ifx{#1\undefined}
}%
\providecommand \@ifnum [1]{%
 \ifnum #1\expandafter \@firstoftwo
 \else \expandafter \@secondoftwo
 \fi
}%
\providecommand \@ifx [1]{%
 \ifx #1\expandafter \@firstoftwo
 \else \expandafter \@secondoftwo
 \fi
}%
\providecommand \natexlab [1]{#1}%
\providecommand \enquote  [1]{``#1''}%
\providecommand \bibnamefont  [1]{#1}%
\providecommand \bibfnamefont [1]{#1}%
\providecommand \citenamefont [1]{#1}%
\providecommand \href@noop [0]{\@secondoftwo}%
\providecommand \href [0]{\begingroup \@sanitize@url \@href}%
\providecommand \@href[1]{\@@startlink{#1}\@@href}%
\providecommand \@@href[1]{\endgroup#1\@@endlink}%
\providecommand \@sanitize@url [0]{\catcode `\\12\catcode `\$12\catcode
  `\&12\catcode `\#12\catcode `\^12\catcode `\_12\catcode `\%12\relax}%
\providecommand \@@startlink[1]{}%
\providecommand \@@endlink[0]{}%
\providecommand \url  [0]{\begingroup\@sanitize@url \@url }%
\providecommand \@url [1]{\endgroup\@href {#1}{\urlprefix }}%
\providecommand \urlprefix  [0]{URL }%
\providecommand \Eprint [0]{\href }%
\providecommand \doibase [0]{http://dx.doi.org/}%
\providecommand \selectlanguage [0]{\@gobble}%
\providecommand \bibinfo  [0]{\@secondoftwo}%
\providecommand \bibfield  [0]{\@secondoftwo}%
\providecommand \translation [1]{[#1]}%
\providecommand \BibitemOpen [0]{}%
\providecommand \bibitemStop [0]{}%
\providecommand \bibitemNoStop [0]{.\EOS\space}%
\providecommand \EOS [0]{\spacefactor3000\relax}%
\providecommand \BibitemShut  [1]{\csname bibitem#1\endcsname}%
\let\auto@bib@innerbib\@empty
\bibitem [{\citenamefont {Shlesinger}\ \emph {et~al.}(1982)\citenamefont
  {Shlesinger}, \citenamefont {Klafter},\ and\ \citenamefont
  {Wong}}]{shlesinger1982random}%
  \BibitemOpen
  \bibfield  {author} {\bibinfo {author} {\bibfnamefont {M.~F.}\ \bibnamefont
  {Shlesinger}}, \bibinfo {author} {\bibfnamefont {J.}~\bibnamefont {Klafter}},
  \ and\ \bibinfo {author} {\bibfnamefont {Y.~M.}\ \bibnamefont {Wong}},\
  }\href@noop {} {\bibfield  {journal} {\bibinfo  {journal} {Journal of
  Statistical Physics}\ }\textbf {\bibinfo {volume} {27}},\ \bibinfo {pages}
  {499} (\bibinfo {year} {1982})}\BibitemShut {NoStop}%
\bibitem [{\citenamefont {Shlesinger}\ and\ \citenamefont
  {Klafter}(1986)}]{shlesinger1986levy}%
  \BibitemOpen
  \bibfield  {author} {\bibinfo {author} {\bibfnamefont {M.~F.}\ \bibnamefont
  {Shlesinger}}\ and\ \bibinfo {author} {\bibfnamefont {J.}~\bibnamefont
  {Klafter}},\ }in\ \href@noop {} {\emph {\bibinfo {booktitle} {On growth and
  form}}}\ (\bibinfo  {publisher} {Springer},\ \bibinfo {year} {1986})\ pp.\
  \bibinfo {pages} {279--283}\BibitemShut {NoStop}%
\bibitem [{\citenamefont {Shlesinger}\ \emph {et~al.}(1987)\citenamefont
  {Shlesinger}, \citenamefont {West},\ and\ \citenamefont
  {Klafter}}]{shlesinger1987levy}%
  \BibitemOpen
  \bibfield  {author} {\bibinfo {author} {\bibfnamefont {M.~F.}\ \bibnamefont
  {Shlesinger}}, \bibinfo {author} {\bibfnamefont {B.~J.}\ \bibnamefont
  {West}}, \ and\ \bibinfo {author} {\bibfnamefont {J.}~\bibnamefont
  {Klafter}},\ }\href@noop {} {\bibfield  {journal} {\bibinfo  {journal}
  {Physical Review Letters}\ }\textbf {\bibinfo {volume} {58}},\ \bibinfo
  {pages} {1100} (\bibinfo {year} {1987})}\BibitemShut {NoStop}%
\bibitem [{\citenamefont {Mandelbrot}\ and\ \citenamefont
  {Pignoni}(1983)}]{mandelbrot1983fractal}%
  \BibitemOpen
  \bibfield  {author} {\bibinfo {author} {\bibfnamefont {B.~B.}\ \bibnamefont
  {Mandelbrot}}\ and\ \bibinfo {author} {\bibfnamefont {R.}~\bibnamefont
  {Pignoni}},\ }\href@noop {} {\  (\bibinfo {year} {1983})}\BibitemShut
  {NoStop}%
\bibitem [{\citenamefont {Metzler}\ and\ \citenamefont
  {Klafter}(2000)}]{metzler2000random}%
  \BibitemOpen
  \bibfield  {author} {\bibinfo {author} {\bibfnamefont {R.}~\bibnamefont
  {Metzler}}\ and\ \bibinfo {author} {\bibfnamefont {J.}~\bibnamefont
  {Klafter}},\ }\href@noop {} {\bibfield  {journal} {\bibinfo  {journal}
  {Physics Reports}\ }\textbf {\bibinfo {volume} {339}},\ \bibinfo {pages} {1}
  (\bibinfo {year} {2000})}\BibitemShut {NoStop}%
\bibitem [{\citenamefont {Harris}\ \emph {et~al.}(2012)\citenamefont {Harris},
  \citenamefont {Banigan}, \citenamefont {Christian}, \citenamefont {Konradt},
  \citenamefont {Wojno}, \citenamefont {Norose}, \citenamefont {Wilson},
  \citenamefont {John}, \citenamefont {Weninger}, \citenamefont {Luster} \emph
  {et~al.}}]{harris2012generalized}%
  \BibitemOpen
  \bibfield  {author} {\bibinfo {author} {\bibfnamefont {T.~H.}\ \bibnamefont
  {Harris}}, \bibinfo {author} {\bibfnamefont {E.~J.}\ \bibnamefont {Banigan}},
  \bibinfo {author} {\bibfnamefont {D.~A.}\ \bibnamefont {Christian}}, \bibinfo
  {author} {\bibfnamefont {C.}~\bibnamefont {Konradt}}, \bibinfo {author}
  {\bibfnamefont {E.~D.~T.}\ \bibnamefont {Wojno}}, \bibinfo {author}
  {\bibfnamefont {K.}~\bibnamefont {Norose}}, \bibinfo {author} {\bibfnamefont
  {E.~H.}\ \bibnamefont {Wilson}}, \bibinfo {author} {\bibfnamefont
  {B.}~\bibnamefont {John}}, \bibinfo {author} {\bibfnamefont {W.}~\bibnamefont
  {Weninger}}, \bibinfo {author} {\bibfnamefont {A.~D.}\ \bibnamefont
  {Luster}},  \emph {et~al.},\ }\href@noop {} {\bibfield  {journal} {\bibinfo
  {journal} {Nature}\ }\textbf {\bibinfo {volume} {486}},\ \bibinfo {pages}
  {545} (\bibinfo {year} {2012})}\BibitemShut {NoStop}%
\bibitem [{\citenamefont {Grinberg}\ \emph {et~al.}(2011)\citenamefont
  {Grinberg}, \citenamefont {Milton},\ and\ \citenamefont
  {Kraig}}]{grinberg2011spreading}%
  \BibitemOpen
  \bibfield  {author} {\bibinfo {author} {\bibfnamefont {Y.~Y.}\ \bibnamefont
  {Grinberg}}, \bibinfo {author} {\bibfnamefont {J.~G.}\ \bibnamefont
  {Milton}}, \ and\ \bibinfo {author} {\bibfnamefont {R.~P.}\ \bibnamefont
  {Kraig}},\ }\href@noop {} {\bibfield  {journal} {\bibinfo  {journal} {PloS
  one}\ }\textbf {\bibinfo {volume} {6}},\ \bibinfo {pages} {e19294} (\bibinfo
  {year} {2011})}\BibitemShut {NoStop}%
\bibitem [{\citenamefont {Takagi}\ \emph {et~al.}(2008)\citenamefont {Takagi},
  \citenamefont {Sato}, \citenamefont {Yanagida},\ and\ \citenamefont
  {Ueda}}]{takagi2008functional}%
  \BibitemOpen
  \bibfield  {author} {\bibinfo {author} {\bibfnamefont {H.}~\bibnamefont
  {Takagi}}, \bibinfo {author} {\bibfnamefont {M.~J.}\ \bibnamefont {Sato}},
  \bibinfo {author} {\bibfnamefont {T.}~\bibnamefont {Yanagida}}, \ and\
  \bibinfo {author} {\bibfnamefont {M.}~\bibnamefont {Ueda}},\ }\href@noop {}
  {\bibfield  {journal} {\bibinfo  {journal} {PloS one}\ }\textbf {\bibinfo
  {volume} {3}},\ \bibinfo {pages} {e2648} (\bibinfo {year}
  {2008})}\BibitemShut {NoStop}%
\bibitem [{\citenamefont {Reynolds}(2010)}]{reynolds2010can}%
  \BibitemOpen
  \bibfield  {author} {\bibinfo {author} {\bibfnamefont {A.~M.}\ \bibnamefont
  {Reynolds}},\ }\href@noop {} {\bibfield  {journal} {\bibinfo  {journal}
  {Physica A: Statistical Mechanics and its Applications}\ }\textbf {\bibinfo
  {volume} {389}},\ \bibinfo {pages} {273} (\bibinfo {year}
  {2010})}\BibitemShut {NoStop}%
\bibitem [{\citenamefont {Ariel}\ \emph {et~al.}(2015)\citenamefont {Ariel},
  \citenamefont {Rabani}, \citenamefont {Benisty}, \citenamefont {Partridge},
  \citenamefont {Harshey},\ and\ \citenamefont {Be'Er}}]{ariel2015swarming}%
  \BibitemOpen
  \bibfield  {author} {\bibinfo {author} {\bibfnamefont {G.}~\bibnamefont
  {Ariel}}, \bibinfo {author} {\bibfnamefont {A.}~\bibnamefont {Rabani}},
  \bibinfo {author} {\bibfnamefont {S.}~\bibnamefont {Benisty}}, \bibinfo
  {author} {\bibfnamefont {J.~D.}\ \bibnamefont {Partridge}}, \bibinfo {author}
  {\bibfnamefont {R.~M.}\ \bibnamefont {Harshey}}, \ and\ \bibinfo {author}
  {\bibfnamefont {A.}~\bibnamefont {Be'Er}},\ }\href@noop {} {\bibfield
  {journal} {\bibinfo  {journal} {Nature communications}\ }\textbf {\bibinfo
  {volume} {6}} (\bibinfo {year} {2015})}\BibitemShut {NoStop}%
\bibitem [{\citenamefont {Reynolds}\ and\ \citenamefont
  {Frye}(2007)}]{reynolds2007free}%
  \BibitemOpen
  \bibfield  {author} {\bibinfo {author} {\bibfnamefont {A.~M.}\ \bibnamefont
  {Reynolds}}\ and\ \bibinfo {author} {\bibfnamefont {M.~A.}\ \bibnamefont
  {Frye}},\ }\href@noop {} {\bibfield  {journal} {\bibinfo  {journal} {PloS
  one}\ }\textbf {\bibinfo {volume} {2}},\ \bibinfo {pages} {e354} (\bibinfo
  {year} {2007})}\BibitemShut {NoStop}%
\bibitem [{\citenamefont {Reynolds}\ \emph {et~al.}(2007)\citenamefont
  {Reynolds}, \citenamefont {Smith}, \citenamefont {Menzel}, \citenamefont
  {Greggers}, \citenamefont {Reynolds},\ and\ \citenamefont
  {Riley}}]{reynolds2007displaced}%
  \BibitemOpen
  \bibfield  {author} {\bibinfo {author} {\bibfnamefont {A.~M.}\ \bibnamefont
  {Reynolds}}, \bibinfo {author} {\bibfnamefont {A.~D.}\ \bibnamefont {Smith}},
  \bibinfo {author} {\bibfnamefont {R.}~\bibnamefont {Menzel}}, \bibinfo
  {author} {\bibfnamefont {U.}~\bibnamefont {Greggers}}, \bibinfo {author}
  {\bibfnamefont {D.~R.}\ \bibnamefont {Reynolds}}, \ and\ \bibinfo {author}
  {\bibfnamefont {J.~R.}\ \bibnamefont {Riley}},\ }\href@noop {} {\bibfield
  {journal} {\bibinfo  {journal} {Ecology}\ }\textbf {\bibinfo {volume} {88}},\
  \bibinfo {pages} {1955} (\bibinfo {year} {2007})}\BibitemShut {NoStop}%
\bibitem [{\citenamefont {Wolf}\ \emph {et~al.}(2016)\citenamefont {Wolf},
  \citenamefont {Nicholls}, \citenamefont {Reynolds}, \citenamefont {Wells},
  \citenamefont {Lim}, \citenamefont {Paxton},\ and\ \citenamefont
  {Osborne}}]{wolf2016optimal}%
  \BibitemOpen
  \bibfield  {author} {\bibinfo {author} {\bibfnamefont {S.}~\bibnamefont
  {Wolf}}, \bibinfo {author} {\bibfnamefont {E.}~\bibnamefont {Nicholls}},
  \bibinfo {author} {\bibfnamefont {A.~M.}\ \bibnamefont {Reynolds}}, \bibinfo
  {author} {\bibfnamefont {P.}~\bibnamefont {Wells}}, \bibinfo {author}
  {\bibfnamefont {K.~S.}\ \bibnamefont {Lim}}, \bibinfo {author} {\bibfnamefont
  {R.~J.}\ \bibnamefont {Paxton}}, \ and\ \bibinfo {author} {\bibfnamefont
  {J.~L.}\ \bibnamefont {Osborne}},\ }\href@noop {} {\bibfield  {journal}
  {\bibinfo  {journal} {Scientific Reports}\ }\textbf {\bibinfo {volume} {6}}
  (\bibinfo {year} {2016})}\BibitemShut {NoStop}%
\bibitem [{\citenamefont {Viswanathan}\ \emph {et~al.}(1996)\citenamefont
  {Viswanathan}, \citenamefont {Afanasyev}, \citenamefont {Buldyrev},
  \citenamefont {Murphy} \emph {et~al.}}]{viswanathan1996levy}%
  \BibitemOpen
  \bibfield  {author} {\bibinfo {author} {\bibfnamefont {G.~M.}\ \bibnamefont
  {Viswanathan}}, \bibinfo {author} {\bibfnamefont {V.}~\bibnamefont
  {Afanasyev}}, \bibinfo {author} {\bibfnamefont {S.~V.}\ \bibnamefont
  {Buldyrev}}, \bibinfo {author} {\bibfnamefont {E.~J.}\ \bibnamefont
  {Murphy}},  \emph {et~al.},\ }\href@noop {} {\bibfield  {journal} {\bibinfo
  {journal} {Nature}\ }\textbf {\bibinfo {volume} {381}},\ \bibinfo {pages}
  {413} (\bibinfo {year} {1996})}\BibitemShut {NoStop}%
\bibitem [{\citenamefont {Humphries}\ \emph {et~al.}(2013)\citenamefont
  {Humphries}, \citenamefont {Weimerskirch},\ and\ \citenamefont
  {Sims}}]{humphries2013new}%
  \BibitemOpen
  \bibfield  {author} {\bibinfo {author} {\bibfnamefont {N.~E.}\ \bibnamefont
  {Humphries}}, \bibinfo {author} {\bibfnamefont {H.}~\bibnamefont
  {Weimerskirch}}, \ and\ \bibinfo {author} {\bibfnamefont {D.~W.}\
  \bibnamefont {Sims}},\ }\href@noop {} {\bibfield  {journal} {\bibinfo
  {journal} {Methods in Ecology and Evolution}\ }\textbf {\bibinfo {volume}
  {4}},\ \bibinfo {pages} {930} (\bibinfo {year} {2013})}\BibitemShut {NoStop}%
\bibitem [{\citenamefont {Sims}\ \emph {et~al.}(2008)\citenamefont {Sims},
  \citenamefont {Southall}, \citenamefont {Humphries}, \citenamefont {Hays},
  \citenamefont {Bradshaw}, \citenamefont {Pitchford}, \citenamefont {James},
  \citenamefont {Ahmed}, \citenamefont {Brierley}, \citenamefont {Hindell}
  \emph {et~al.}}]{sims2008scaling}%
  \BibitemOpen
  \bibfield  {author} {\bibinfo {author} {\bibfnamefont {D.~W.}\ \bibnamefont
  {Sims}}, \bibinfo {author} {\bibfnamefont {E.~J.}\ \bibnamefont {Southall}},
  \bibinfo {author} {\bibfnamefont {N.~E.}\ \bibnamefont {Humphries}}, \bibinfo
  {author} {\bibfnamefont {G.~C.}\ \bibnamefont {Hays}}, \bibinfo {author}
  {\bibfnamefont {C.~J.~A.}\ \bibnamefont {Bradshaw}}, \bibinfo {author}
  {\bibfnamefont {J.~W.}\ \bibnamefont {Pitchford}}, \bibinfo {author}
  {\bibfnamefont {A.}~\bibnamefont {James}}, \bibinfo {author} {\bibfnamefont
  {M.~Z.}\ \bibnamefont {Ahmed}}, \bibinfo {author} {\bibfnamefont {A.~S.}\
  \bibnamefont {Brierley}}, \bibinfo {author} {\bibfnamefont {M.~A.}\
  \bibnamefont {Hindell}},  \emph {et~al.},\ }\href@noop {} {\bibfield
  {journal} {\bibinfo  {journal} {Nature}\ }\textbf {\bibinfo {volume} {451}},\
  \bibinfo {pages} {1098} (\bibinfo {year} {2008})}\BibitemShut {NoStop}%
\bibitem [{\citenamefont {Brockmann}\ \emph {et~al.}(2006)\citenamefont
  {Brockmann}, \citenamefont {Hufnagel},\ and\ \citenamefont
  {Geisel}}]{brockmann2006scaling}%
  \BibitemOpen
  \bibfield  {author} {\bibinfo {author} {\bibfnamefont {D.}~\bibnamefont
  {Brockmann}}, \bibinfo {author} {\bibfnamefont {L.}~\bibnamefont {Hufnagel}},
  \ and\ \bibinfo {author} {\bibfnamefont {T.}~\bibnamefont {Geisel}},\
  }\href@noop {} {\bibfield  {journal} {\bibinfo  {journal} {Nature}\ }\textbf
  {\bibinfo {volume} {439}},\ \bibinfo {pages} {462} (\bibinfo {year}
  {2006})}\BibitemShut {NoStop}%
\bibitem [{\citenamefont {Gonzalez}\ \emph {et~al.}(2008)\citenamefont
  {Gonzalez}, \citenamefont {Hidalgo},\ and\ \citenamefont
  {Barabasi}}]{gonzalez2008understanding}%
  \BibitemOpen
  \bibfield  {author} {\bibinfo {author} {\bibfnamefont {M.~C.}\ \bibnamefont
  {Gonzalez}}, \bibinfo {author} {\bibfnamefont {C.~A.}\ \bibnamefont
  {Hidalgo}}, \ and\ \bibinfo {author} {\bibfnamefont {A.-L.}\ \bibnamefont
  {Barabasi}},\ }\href@noop {} {\bibfield  {journal} {\bibinfo  {journal}
  {Nature}\ }\textbf {\bibinfo {volume} {453}},\ \bibinfo {pages} {779}
  (\bibinfo {year} {2008})}\BibitemShut {NoStop}%
\bibitem [{\citenamefont {Rhee}\ \emph {et~al.}(2011)\citenamefont {Rhee},
  \citenamefont {Shin}, \citenamefont {Hong}, \citenamefont {Lee},
  \citenamefont {Kim},\ and\ \citenamefont {Chong}}]{rhee2011levy}%
  \BibitemOpen
  \bibfield  {author} {\bibinfo {author} {\bibfnamefont {I.}~\bibnamefont
  {Rhee}}, \bibinfo {author} {\bibfnamefont {M.}~\bibnamefont {Shin}}, \bibinfo
  {author} {\bibfnamefont {S.}~\bibnamefont {Hong}}, \bibinfo {author}
  {\bibfnamefont {K.}~\bibnamefont {Lee}}, \bibinfo {author} {\bibfnamefont
  {S.~J.}\ \bibnamefont {Kim}}, \ and\ \bibinfo {author} {\bibfnamefont
  {S.}~\bibnamefont {Chong}},\ }\href@noop {} {\bibfield  {journal} {\bibinfo
  {journal} {IEEE/ACM transactions on networking (TON)}\ }\textbf {\bibinfo
  {volume} {19}},\ \bibinfo {pages} {630} (\bibinfo {year} {2011})}\BibitemShut
  {NoStop}%
\bibitem [{\citenamefont {Brockmann}\ and\ \citenamefont
  {Geisel}(2000)}]{brockmann2000ecology}%
  \BibitemOpen
  \bibfield  {author} {\bibinfo {author} {\bibfnamefont {D.}~\bibnamefont
  {Brockmann}}\ and\ \bibinfo {author} {\bibfnamefont {T.}~\bibnamefont
  {Geisel}},\ }\href@noop {} {\bibfield  {journal} {\bibinfo  {journal}
  {Neurocomputing}\ }\textbf {\bibinfo {volume} {32}},\ \bibinfo {pages} {643}
  (\bibinfo {year} {2000})}\BibitemShut {NoStop}%
\bibitem [{\citenamefont {Costa}\ \emph {et~al.}(2009)\citenamefont {Costa},
  \citenamefont {Bonomo},\ and\ \citenamefont {Sigman}}]{costa2009scale}%
  \BibitemOpen
  \bibfield  {author} {\bibinfo {author} {\bibfnamefont {M.~E.}\ \bibnamefont
  {Costa}}, \bibinfo {author} {\bibfnamefont {F.}~\bibnamefont {Bonomo}}, \
  and\ \bibinfo {author} {\bibfnamefont {M.}~\bibnamefont {Sigman}},\
  }\href@noop {} {\bibfield  {journal} {\bibinfo  {journal} {Frontiers in
  integrative neuroscience}\ }\textbf {\bibinfo {volume} {3}} (\bibinfo {year}
  {2009})}\BibitemShut {NoStop}%
\bibitem [{\citenamefont {Viswanathan}\ \emph {et~al.}(1999)\citenamefont
  {Viswanathan}, \citenamefont {Buldyrev}, \citenamefont {Havlin},
  \citenamefont {Da~Luz}, \citenamefont {Raposo},\ and\ \citenamefont
  {Stanley}}]{viswanathan1999optimizing}%
  \BibitemOpen
  \bibfield  {author} {\bibinfo {author} {\bibfnamefont {G.~M.}\ \bibnamefont
  {Viswanathan}}, \bibinfo {author} {\bibfnamefont {S.~V.}\ \bibnamefont
  {Buldyrev}}, \bibinfo {author} {\bibfnamefont {S.}~\bibnamefont {Havlin}},
  \bibinfo {author} {\bibfnamefont {M.~G.~E.}\ \bibnamefont {Da~Luz}}, \bibinfo
  {author} {\bibfnamefont {E.~P.}\ \bibnamefont {Raposo}}, \ and\ \bibinfo
  {author} {\bibfnamefont {H.~E.}\ \bibnamefont {Stanley}},\ }\href@noop {}
  {\bibfield  {journal} {\bibinfo  {journal} {Nature}\ }\textbf {\bibinfo
  {volume} {401}},\ \bibinfo {pages} {911} (\bibinfo {year}
  {1999})}\BibitemShut {NoStop}%
\bibitem [{\citenamefont {Buldyrev}\ \emph {et~al.}(2001)\citenamefont
  {Buldyrev}, \citenamefont {Havlin}, \citenamefont {Kazakov}, \citenamefont
  {Da~Luz}, \citenamefont {Raposo}, \citenamefont {Stanley},\ and\
  \citenamefont {Viswanathan}}]{buldyrev2001average}%
  \BibitemOpen
  \bibfield  {author} {\bibinfo {author} {\bibfnamefont {S.~V.}\ \bibnamefont
  {Buldyrev}}, \bibinfo {author} {\bibfnamefont {S.}~\bibnamefont {Havlin}},
  \bibinfo {author} {\bibfnamefont {A.~Y.}\ \bibnamefont {Kazakov}}, \bibinfo
  {author} {\bibfnamefont {M.~G.~E.}\ \bibnamefont {Da~Luz}}, \bibinfo {author}
  {\bibfnamefont {E.~P.}\ \bibnamefont {Raposo}}, \bibinfo {author}
  {\bibfnamefont {H.~E.}\ \bibnamefont {Stanley}}, \ and\ \bibinfo {author}
  {\bibfnamefont {G.~M.}\ \bibnamefont {Viswanathan}},\ }\href@noop {}
  {\bibfield  {journal} {\bibinfo  {journal} {Physical Review E}\ }\textbf
  {\bibinfo {volume} {64}},\ \bibinfo {pages} {041108} (\bibinfo {year}
  {2001})}\BibitemShut {NoStop}%
\bibitem [{\citenamefont {Bartumeus}\ \emph {et~al.}(2002)\citenamefont
  {Bartumeus}, \citenamefont {Catalan}, \citenamefont {Fulco}, \citenamefont
  {Lyra},\ and\ \citenamefont {Viswanathan}}]{bartumeus2002optimizing}%
  \BibitemOpen
  \bibfield  {author} {\bibinfo {author} {\bibfnamefont {F.}~\bibnamefont
  {Bartumeus}}, \bibinfo {author} {\bibfnamefont {J.}~\bibnamefont {Catalan}},
  \bibinfo {author} {\bibfnamefont {U.~L.}\ \bibnamefont {Fulco}}, \bibinfo
  {author} {\bibfnamefont {M.~L.}\ \bibnamefont {Lyra}}, \ and\ \bibinfo
  {author} {\bibfnamefont {G.~M.}\ \bibnamefont {Viswanathan}},\ }\href@noop {}
  {\bibfield  {journal} {\bibinfo  {journal} {Phys. Rev. Lett.}\ }\textbf
  {\bibinfo {volume} {88}},\ \bibinfo {pages} {097901} (\bibinfo {year}
  {2002})}\BibitemShut {NoStop}%
\bibitem [{\citenamefont {Raposo}\ \emph {et~al.}(2003)\citenamefont {Raposo},
  \citenamefont {Buldyrev}, \citenamefont {Da~Luz}, \citenamefont {Santos},
  \citenamefont {Stanley},\ and\ \citenamefont
  {Viswanathan}}]{raposo2003dynamical}%
  \BibitemOpen
  \bibfield  {author} {\bibinfo {author} {\bibfnamefont {E.~P.}\ \bibnamefont
  {Raposo}}, \bibinfo {author} {\bibfnamefont {S.~V.}\ \bibnamefont
  {Buldyrev}}, \bibinfo {author} {\bibfnamefont {M.~G.~E.}\ \bibnamefont
  {Da~Luz}}, \bibinfo {author} {\bibfnamefont {M.~C.}\ \bibnamefont {Santos}},
  \bibinfo {author} {\bibfnamefont {H.~E.}\ \bibnamefont {Stanley}}, \ and\
  \bibinfo {author} {\bibfnamefont {G.~M.}\ \bibnamefont {Viswanathan}},\
  }\href@noop {} {\bibfield  {journal} {\bibinfo  {journal} {Physical review
  letters}\ }\textbf {\bibinfo {volume} {91}},\ \bibinfo {pages} {240601}
  (\bibinfo {year} {2003})}\BibitemShut {NoStop}%
\bibitem [{\citenamefont {Bartumeus}\ \emph {et~al.}(2005)\citenamefont
  {Bartumeus}, \citenamefont {da~Luz}, \citenamefont {Viswanathan},\ and\
  \citenamefont {Catalan}}]{bartumeus2005animal}%
  \BibitemOpen
  \bibfield  {author} {\bibinfo {author} {\bibfnamefont {F.}~\bibnamefont
  {Bartumeus}}, \bibinfo {author} {\bibfnamefont {M.~G.~E.}\ \bibnamefont
  {da~Luz}}, \bibinfo {author} {\bibfnamefont {G.~M.}\ \bibnamefont
  {Viswanathan}}, \ and\ \bibinfo {author} {\bibfnamefont {J.}~\bibnamefont
  {Catalan}},\ }\href@noop {} {\bibfield  {journal} {\bibinfo  {journal}
  {Ecology}\ }\textbf {\bibinfo {volume} {86}},\ \bibinfo {pages} {3078}
  (\bibinfo {year} {2005})}\BibitemShut {NoStop}%
\bibitem [{\citenamefont {Pasternak}\ \emph {et~al.}(2009)\citenamefont
  {Pasternak}, \citenamefont {Bartumeus},\ and\ \citenamefont
  {Grasso}}]{pasternak2009levy}%
  \BibitemOpen
  \bibfield  {author} {\bibinfo {author} {\bibfnamefont {Z.}~\bibnamefont
  {Pasternak}}, \bibinfo {author} {\bibfnamefont {F.}~\bibnamefont
  {Bartumeus}}, \ and\ \bibinfo {author} {\bibfnamefont {F.~W.}\ \bibnamefont
  {Grasso}},\ }\href@noop {} {\bibfield  {journal} {\bibinfo  {journal}
  {Journal of Physics A: Mathematical and Theoretical}\ }\textbf {\bibinfo
  {volume} {42}},\ \bibinfo {pages} {434010} (\bibinfo {year}
  {2009})}\BibitemShut {NoStop}%
\bibitem [{\citenamefont {Raposo}\ \emph {et~al.}(2011)\citenamefont {Raposo},
  \citenamefont {Bartumeus}, \citenamefont {Da~Luz}, \citenamefont
  {Ribeiro-Neto}, \citenamefont {Souza},\ and\ \citenamefont
  {Viswanathan}}]{raposo2011landscape}%
  \BibitemOpen
  \bibfield  {author} {\bibinfo {author} {\bibfnamefont {E.~P.}\ \bibnamefont
  {Raposo}}, \bibinfo {author} {\bibfnamefont {F.}~\bibnamefont {Bartumeus}},
  \bibinfo {author} {\bibfnamefont {M.~G.~E.}\ \bibnamefont {Da~Luz}}, \bibinfo
  {author} {\bibfnamefont {P.~J.}\ \bibnamefont {Ribeiro-Neto}}, \bibinfo
  {author} {\bibfnamefont {T.~A.}\ \bibnamefont {Souza}}, \ and\ \bibinfo
  {author} {\bibfnamefont {G.~M.}\ \bibnamefont {Viswanathan}},\ }\href@noop {}
  {\bibfield  {journal} {\bibinfo  {journal} {PLoS Comput Biol}\ }\textbf
  {\bibinfo {volume} {7}},\ \bibinfo {pages} {e1002233} (\bibinfo {year}
  {2011})}\BibitemShut {NoStop}%
\bibitem [{\citenamefont {Palyulin}\ \emph {et~al.}(2014)\citenamefont
  {Palyulin}, \citenamefont {Chechkin},\ and\ \citenamefont
  {Metzler}}]{palyulin2014levy}%
  \BibitemOpen
  \bibfield  {author} {\bibinfo {author} {\bibfnamefont {V.~V.}\ \bibnamefont
  {Palyulin}}, \bibinfo {author} {\bibfnamefont {A.~V.}\ \bibnamefont
  {Chechkin}}, \ and\ \bibinfo {author} {\bibfnamefont {R.}~\bibnamefont
  {Metzler}},\ }\href@noop {} {\bibfield  {journal} {\bibinfo  {journal}
  {Proceedings of the National Academy of Sciences}\ }\textbf {\bibinfo
  {volume} {111}},\ \bibinfo {pages} {2931} (\bibinfo {year}
  {2014})}\BibitemShut {NoStop}%
\bibitem [{\citenamefont {Humphries}\ and\ \citenamefont
  {Sims}(2014)}]{humphries2014optimal}%
  \BibitemOpen
  \bibfield  {author} {\bibinfo {author} {\bibfnamefont {N.~E.}\ \bibnamefont
  {Humphries}}\ and\ \bibinfo {author} {\bibfnamefont {D.~W.}\ \bibnamefont
  {Sims}},\ }\href@noop {} {\bibfield  {journal} {\bibinfo  {journal} {Journal
  of theoretical biology}\ }\textbf {\bibinfo {volume} {358}},\ \bibinfo
  {pages} {179} (\bibinfo {year} {2014})}\BibitemShut {NoStop}%
\bibitem [{\citenamefont {Kusmierz}\ \emph {et~al.}(2014)\citenamefont
  {Kusmierz}, \citenamefont {Majumdar}, \citenamefont {Sabhapandit},\ and\
  \citenamefont {Schehr}}]{kusmierz2014first}%
  \BibitemOpen
  \bibfield  {author} {\bibinfo {author} {\bibfnamefont {L.}~\bibnamefont
  {Kusmierz}}, \bibinfo {author} {\bibfnamefont {S.~N.}\ \bibnamefont
  {Majumdar}}, \bibinfo {author} {\bibfnamefont {S.}~\bibnamefont
  {Sabhapandit}}, \ and\ \bibinfo {author} {\bibfnamefont {G.}~\bibnamefont
  {Schehr}},\ }\href@noop {} {\bibfield  {journal} {\bibinfo  {journal}
  {Physical Review Letters}\ }\textbf {\bibinfo {volume} {113}},\ \bibinfo
  {pages} {220602} (\bibinfo {year} {2014})}\BibitemShut {NoStop}%
\bibitem [{\citenamefont {Ku{\'s}mierz}\ and\ \citenamefont
  {Gudowska-Nowak}(2015)}]{kusmierz2015optimal}%
  \BibitemOpen
  \bibfield  {author} {\bibinfo {author} {\bibfnamefont {{\L}.}~\bibnamefont
  {Ku{\'s}mierz}}\ and\ \bibinfo {author} {\bibfnamefont {E.}~\bibnamefont
  {Gudowska-Nowak}},\ }\href@noop {} {\bibfield  {journal} {\bibinfo  {journal}
  {Physical Review E}\ }\textbf {\bibinfo {volume} {92}},\ \bibinfo {pages}
  {052127} (\bibinfo {year} {2015})}\BibitemShut {NoStop}%
\bibitem [{\citenamefont {Reynolds}(2009)}]{reynolds2009scale}%
  \BibitemOpen
  \bibfield  {author} {\bibinfo {author} {\bibfnamefont {A.~M.}\ \bibnamefont
  {Reynolds}},\ }\href@noop {} {\bibfield  {journal} {\bibinfo  {journal}
  {Journal of Physics A: Mathematical and Theoretical}\ }\textbf {\bibinfo
  {volume} {42}},\ \bibinfo {pages} {434006} (\bibinfo {year}
  {2009})}\BibitemShut {NoStop}%
\bibitem [{\citenamefont {Bartumeus}(2007)}]{bartumeus2007levy}%
  \BibitemOpen
  \bibfield  {author} {\bibinfo {author} {\bibfnamefont {F.}~\bibnamefont
  {Bartumeus}},\ }\href@noop {} {\bibfield  {journal} {\bibinfo  {journal}
  {Fractals}\ }\textbf {\bibinfo {volume} {15}},\ \bibinfo {pages} {151}
  (\bibinfo {year} {2007})}\BibitemShut {NoStop}%
\bibitem [{\citenamefont {Viswanathan}\ \emph {et~al.}(2008)\citenamefont
  {Viswanathan}, \citenamefont {Raposo},\ and\ \citenamefont
  {Da~Luz}}]{viswanathan2008levy}%
  \BibitemOpen
  \bibfield  {author} {\bibinfo {author} {\bibfnamefont {G.~M.}\ \bibnamefont
  {Viswanathan}}, \bibinfo {author} {\bibfnamefont {E.~P.}\ \bibnamefont
  {Raposo}}, \ and\ \bibinfo {author} {\bibfnamefont {M.~G.~E.}\ \bibnamefont
  {Da~Luz}},\ }\href@noop {} {\bibfield  {journal} {\bibinfo  {journal}
  {Physics of Life Reviews}\ }\textbf {\bibinfo {volume} {5}},\ \bibinfo
  {pages} {133} (\bibinfo {year} {2008})}\BibitemShut {NoStop}%
\bibitem [{\citenamefont {Taub}\ \emph {et~al.}(1993)\citenamefont {Taub},
  \citenamefont {Conlon}, \citenamefont {Lloyd}, \citenamefont {Oppenheim},\
  and\ \citenamefont {Kelvin}}]{taub1993preferential}%
  \BibitemOpen
  \bibfield  {author} {\bibinfo {author} {\bibfnamefont {D.~D.}\ \bibnamefont
  {Taub}}, \bibinfo {author} {\bibfnamefont {K.}~\bibnamefont {Conlon}},
  \bibinfo {author} {\bibfnamefont {A.~R.}\ \bibnamefont {Lloyd}}, \bibinfo
  {author} {\bibfnamefont {J.~J.}\ \bibnamefont {Oppenheim}}, \ and\ \bibinfo
  {author} {\bibfnamefont {D.~J.}\ \bibnamefont {Kelvin}},\ }\href@noop {}
  {\bibfield  {journal} {\bibinfo  {journal} {Science}\ }\textbf {\bibinfo
  {volume} {260}},\ \bibinfo {pages} {355} (\bibinfo {year}
  {1993})}\BibitemShut {NoStop}%
\bibitem [{\citenamefont {Gunn}\ \emph {et~al.}(1998)\citenamefont {Gunn},
  \citenamefont {Tangemann}, \citenamefont {Tam}, \citenamefont {Cyster},
  \citenamefont {Rosen},\ and\ \citenamefont {Williams}}]{gunn1998chemokine}%
  \BibitemOpen
  \bibfield  {author} {\bibinfo {author} {\bibfnamefont {M.~D.}\ \bibnamefont
  {Gunn}}, \bibinfo {author} {\bibfnamefont {K.}~\bibnamefont {Tangemann}},
  \bibinfo {author} {\bibfnamefont {C.}~\bibnamefont {Tam}}, \bibinfo {author}
  {\bibfnamefont {J.~G.}\ \bibnamefont {Cyster}}, \bibinfo {author}
  {\bibfnamefont {S.~D.}\ \bibnamefont {Rosen}}, \ and\ \bibinfo {author}
  {\bibfnamefont {L.~T.}\ \bibnamefont {Williams}},\ }\href@noop {} {\bibfield
  {journal} {\bibinfo  {journal} {Proceedings of the National Academy of
  Sciences}\ }\textbf {\bibinfo {volume} {95}},\ \bibinfo {pages} {258}
  (\bibinfo {year} {1998})}\BibitemShut {NoStop}%
\bibitem [{\citenamefont {Adler}(1966)}]{adler1966chemotaxis}%
  \BibitemOpen
  \bibfield  {author} {\bibinfo {author} {\bibfnamefont {J.}~\bibnamefont
  {Adler}},\ }\href@noop {} {\bibfield  {journal} {\bibinfo  {journal}
  {Science}\ }\textbf {\bibinfo {volume} {153}},\ \bibinfo {pages} {708}
  (\bibinfo {year} {1966})}\BibitemShut {NoStop}%
\bibitem [{\citenamefont {Berg}\ and\ \citenamefont
  {Brown}(1974)}]{berg1974chemotaxis}%
  \BibitemOpen
  \bibfield  {author} {\bibinfo {author} {\bibfnamefont {H.~C.}\ \bibnamefont
  {Berg}}\ and\ \bibinfo {author} {\bibfnamefont {D.~A.}\ \bibnamefont
  {Brown}},\ }in\ \href@noop {} {\emph {\bibinfo {booktitle} {Chemotaxis: Its
  Biology and Biochemistry}}}\ (\bibinfo  {publisher} {Karger Publishers},\
  \bibinfo {year} {1974})\ pp.\ \bibinfo {pages} {55--78}\BibitemShut {NoStop}%
\bibitem [{\citenamefont {Szurmant}\ and\ \citenamefont
  {Ordal}(2004)}]{szurmant2004diversity}%
  \BibitemOpen
  \bibfield  {author} {\bibinfo {author} {\bibfnamefont {H.}~\bibnamefont
  {Szurmant}}\ and\ \bibinfo {author} {\bibfnamefont {G.~W.}\ \bibnamefont
  {Ordal}},\ }\href@noop {} {\bibfield  {journal} {\bibinfo  {journal}
  {Microbiology and molecular biology reviews}\ }\textbf {\bibinfo {volume}
  {68}},\ \bibinfo {pages} {301} (\bibinfo {year} {2004})}\BibitemShut
  {NoStop}%
\bibitem [{\citenamefont {Zhu}\ \emph {et~al.}(2009)\citenamefont {Zhu},
  \citenamefont {Nern}, \citenamefont {Zipursky},\ and\ \citenamefont
  {Frye}}]{zhu2009peripheral}%
  \BibitemOpen
  \bibfield  {author} {\bibinfo {author} {\bibfnamefont {Y.}~\bibnamefont
  {Zhu}}, \bibinfo {author} {\bibfnamefont {A.}~\bibnamefont {Nern}}, \bibinfo
  {author} {\bibfnamefont {S.~L.}\ \bibnamefont {Zipursky}}, \ and\ \bibinfo
  {author} {\bibfnamefont {M.~A.}\ \bibnamefont {Frye}},\ }\href@noop {}
  {\bibfield  {journal} {\bibinfo  {journal} {Current Biology}\ }\textbf
  {\bibinfo {volume} {19}},\ \bibinfo {pages} {613} (\bibinfo {year}
  {2009})}\BibitemShut {NoStop}%
\bibitem [{\citenamefont {Miller}\ and\ \citenamefont
  {Keller}(1999)}]{miller1999measuring}%
  \BibitemOpen
  \bibfield  {author} {\bibinfo {author} {\bibfnamefont {M.~S.}\ \bibnamefont
  {Miller}}\ and\ \bibinfo {author} {\bibfnamefont {T.~S.}\ \bibnamefont
  {Keller}},\ }\href@noop {} {\bibfield  {journal} {\bibinfo  {journal}
  {Journal of gravitational physiology: a journal of the International Society
  for Gravitational Physiology}\ }\textbf {\bibinfo {volume} {6}},\ \bibinfo
  {pages} {P99} (\bibinfo {year} {1999})}\BibitemShut {NoStop}%
\bibitem [{\citenamefont {Duistermars}\ \emph {et~al.}(2009)\citenamefont
  {Duistermars}, \citenamefont {Chow},\ and\ \citenamefont
  {Frye}}]{duistermars2009flies}%
  \BibitemOpen
  \bibfield  {author} {\bibinfo {author} {\bibfnamefont {B.~J.}\ \bibnamefont
  {Duistermars}}, \bibinfo {author} {\bibfnamefont {D.~M.}\ \bibnamefont
  {Chow}}, \ and\ \bibinfo {author} {\bibfnamefont {M.~A.}\ \bibnamefont
  {Frye}},\ }\href@noop {} {\bibfield  {journal} {\bibinfo  {journal} {Current
  Biology}\ }\textbf {\bibinfo {volume} {19}},\ \bibinfo {pages} {1301}
  (\bibinfo {year} {2009})}\BibitemShut {NoStop}%
\bibitem [{\citenamefont {Wasserman}\ \emph {et~al.}(2013)\citenamefont
  {Wasserman}, \citenamefont {Salomon},\ and\ \citenamefont
  {Frye}}]{wasserman2013drosophila}%
  \BibitemOpen
  \bibfield  {author} {\bibinfo {author} {\bibfnamefont {S.}~\bibnamefont
  {Wasserman}}, \bibinfo {author} {\bibfnamefont {A.}~\bibnamefont {Salomon}},
  \ and\ \bibinfo {author} {\bibfnamefont {M.~A.}\ \bibnamefont {Frye}},\
  }\href@noop {} {\bibfield  {journal} {\bibinfo  {journal} {Current Biology}\
  }\textbf {\bibinfo {volume} {23}},\ \bibinfo {pages} {301} (\bibinfo {year}
  {2013})}\BibitemShut {NoStop}%
\bibitem [{\citenamefont {Reynolds}(2015)}]{reynolds2015liberating}%
  \BibitemOpen
  \bibfield  {author} {\bibinfo {author} {\bibfnamefont {A.~M.}\ \bibnamefont
  {Reynolds}},\ }\href@noop {} {\bibfield  {journal} {\bibinfo  {journal}
  {Physics of life reviews}\ }\textbf {\bibinfo {volume} {14}},\ \bibinfo
  {pages} {59} (\bibinfo {year} {2015})}\BibitemShut {NoStop}%
\bibitem [{\citenamefont {Boyer}\ \emph {et~al.}(2004)\citenamefont {Boyer},
  \citenamefont {Miramontes}, \citenamefont {Ramos-Fernandez}, \citenamefont
  {Mateos},\ and\ \citenamefont {Cocho}}]{boyer2004modeling}%
  \BibitemOpen
  \bibfield  {author} {\bibinfo {author} {\bibfnamefont {D.}~\bibnamefont
  {Boyer}}, \bibinfo {author} {\bibfnamefont {O.}~\bibnamefont {Miramontes}},
  \bibinfo {author} {\bibfnamefont {G.}~\bibnamefont {Ramos-Fernandez}},
  \bibinfo {author} {\bibfnamefont {J.}~\bibnamefont {Mateos}}, \ and\ \bibinfo
  {author} {\bibfnamefont {G.}~\bibnamefont {Cocho}},\ }\href@noop {}
  {\bibfield  {journal} {\bibinfo  {journal} {Physica A: Statistical Mechanics
  and its Applications}\ }\textbf {\bibinfo {volume} {342}},\ \bibinfo {pages}
  {329} (\bibinfo {year} {2004})}\BibitemShut {NoStop}%
\bibitem [{\citenamefont {Santos}\ \emph {et~al.}(2007)\citenamefont {Santos},
  \citenamefont {Boyer}, \citenamefont {Miramontes}, \citenamefont
  {Viswanathan}, \citenamefont {Raposo}, \citenamefont {Mateos},\ and\
  \citenamefont {Da~Luz}}]{santos2007origin}%
  \BibitemOpen
  \bibfield  {author} {\bibinfo {author} {\bibfnamefont {M.~C.}\ \bibnamefont
  {Santos}}, \bibinfo {author} {\bibfnamefont {D.}~\bibnamefont {Boyer}},
  \bibinfo {author} {\bibfnamefont {O.}~\bibnamefont {Miramontes}}, \bibinfo
  {author} {\bibfnamefont {G.~M.}\ \bibnamefont {Viswanathan}}, \bibinfo
  {author} {\bibfnamefont {E.~P.}\ \bibnamefont {Raposo}}, \bibinfo {author}
  {\bibfnamefont {J.~L.}\ \bibnamefont {Mateos}}, \ and\ \bibinfo {author}
  {\bibfnamefont {M.~G.~E.}\ \bibnamefont {Da~Luz}},\ }\href@noop {} {\bibfield
   {journal} {\bibinfo  {journal} {Physical Review E}\ }\textbf {\bibinfo
  {volume} {75}},\ \bibinfo {pages} {061114} (\bibinfo {year}
  {2007})}\BibitemShut {NoStop}%
\bibitem [{\citenamefont {Reynolds}(2008)}]{reynolds2008deterministic}%
  \BibitemOpen
  \bibfield  {author} {\bibinfo {author} {\bibfnamefont {A.~M.}\ \bibnamefont
  {Reynolds}},\ }\href@noop {} {\bibfield  {journal} {\bibinfo  {journal}
  {Physical Review E}\ }\textbf {\bibinfo {volume} {78}},\ \bibinfo {pages}
  {011906} (\bibinfo {year} {2008})}\BibitemShut {NoStop}%
\bibitem [{\citenamefont {Reynolds}(2012)}]{reynolds2012fitness}%
  \BibitemOpen
  \bibfield  {author} {\bibinfo {author} {\bibfnamefont {A.~M.}\ \bibnamefont
  {Reynolds}},\ }\href@noop {} {\bibfield  {journal} {\bibinfo  {journal}
  {Journal of The Royal Society Interface}\ ,\ \bibinfo {pages} {rsif20110815}}
  (\bibinfo {year} {2012})}\BibitemShut {NoStop}%
\bibitem [{\citenamefont {Reynolds}(2014{\natexlab{a}})}]{reynolds2014levy}%
  \BibitemOpen
  \bibfield  {author} {\bibinfo {author} {\bibfnamefont {A.~M.}\ \bibnamefont
  {Reynolds}},\ }in\ \href@noop {} {\emph {\bibinfo {booktitle} {Proc. R. Soc.
  A}}},\ Vol.\ \bibinfo {volume} {470}\ (\bibinfo {organization} {The Royal
  Society},\ \bibinfo {year} {2014})\ p.\ \bibinfo {pages}
  {20140408}\BibitemShut {NoStop}%
\bibitem [{\citenamefont {Shlesinger}(1983)}]{shlesinger1983weierstrassian}%
  \BibitemOpen
  \bibfield  {author} {\bibinfo {author} {\bibfnamefont {M.~F.}\ \bibnamefont
  {Shlesinger}},\ }\href@noop {} {\bibfield  {journal} {\bibinfo  {journal}
  {The Journal of Chemical Physics}\ }\textbf {\bibinfo {volume} {78}},\
  \bibinfo {pages} {416} (\bibinfo {year} {1983})}\BibitemShut {NoStop}%
\bibitem [{\citenamefont {Reynolds}(2014{\natexlab{b}})}]{reynolds2014mussels}%
  \BibitemOpen
  \bibfield  {author} {\bibinfo {author} {\bibfnamefont {A.~M.}\ \bibnamefont
  {Reynolds}},\ }\href@noop {} {\bibfield  {journal} {\bibinfo  {journal}
  {Scientific reports}\ }\textbf {\bibinfo {volume} {4}},\ \bibinfo {pages}
  {4409} (\bibinfo {year} {2014}{\natexlab{b}})}\BibitemShut {NoStop}%
\bibitem [{\citenamefont {Sims}\ \emph {et~al.}(2014)\citenamefont {Sims},
  \citenamefont {Reynolds}, \citenamefont {Humphries}, \citenamefont
  {Southall}, \citenamefont {Wearmouth}, \citenamefont {Metcalfe},\ and\
  \citenamefont {Twitchett}}]{sims2014hierarchical}%
  \BibitemOpen
  \bibfield  {author} {\bibinfo {author} {\bibfnamefont {D.~W.}\ \bibnamefont
  {Sims}}, \bibinfo {author} {\bibfnamefont {A.~M.}\ \bibnamefont {Reynolds}},
  \bibinfo {author} {\bibfnamefont {N.~E.}\ \bibnamefont {Humphries}}, \bibinfo
  {author} {\bibfnamefont {E.~J.}\ \bibnamefont {Southall}}, \bibinfo {author}
  {\bibfnamefont {V.~J.}\ \bibnamefont {Wearmouth}}, \bibinfo {author}
  {\bibfnamefont {B.}~\bibnamefont {Metcalfe}}, \ and\ \bibinfo {author}
  {\bibfnamefont {R.~J.}\ \bibnamefont {Twitchett}},\ }\href@noop {} {\bibfield
   {journal} {\bibinfo  {journal} {Proceedings of the National Academy of
  Sciences}\ }\textbf {\bibinfo {volume} {111}},\ \bibinfo {pages} {11073}
  (\bibinfo {year} {2014})}\BibitemShut {NoStop}%
\bibitem [{\citenamefont {Ott}\ \emph {et~al.}(1990)\citenamefont {Ott},
  \citenamefont {Bouchaud}, \citenamefont {Langevin},\ and\ \citenamefont
  {Urbach}}]{ott1990anomalous}%
  \BibitemOpen
  \bibfield  {author} {\bibinfo {author} {\bibfnamefont {A.}~\bibnamefont
  {Ott}}, \bibinfo {author} {\bibfnamefont {J.~P.}\ \bibnamefont {Bouchaud}},
  \bibinfo {author} {\bibfnamefont {D.}~\bibnamefont {Langevin}}, \ and\
  \bibinfo {author} {\bibfnamefont {W.}~\bibnamefont {Urbach}},\ }\href@noop {}
  {\bibfield  {journal} {\bibinfo  {journal} {Physical review letters}\
  }\textbf {\bibinfo {volume} {65}},\ \bibinfo {pages} {2201} (\bibinfo {year}
  {1990})}\BibitemShut {NoStop}%
\bibitem [{\citenamefont {Srivastava}\ \emph {et~al.}(2009)\citenamefont
  {Srivastava}, \citenamefont {Clark},\ and\ \citenamefont
  {Samuel}}]{srivastava2009temporal}%
  \BibitemOpen
  \bibfield  {author} {\bibinfo {author} {\bibfnamefont {N.}~\bibnamefont
  {Srivastava}}, \bibinfo {author} {\bibfnamefont {D.~A.}\ \bibnamefont
  {Clark}}, \ and\ \bibinfo {author} {\bibfnamefont {A.~D.~T.}\ \bibnamefont
  {Samuel}},\ }\href@noop {} {\bibfield  {journal} {\bibinfo  {journal}
  {Journal of neurophysiology}\ }\textbf {\bibinfo {volume} {102}},\ \bibinfo
  {pages} {1172} (\bibinfo {year} {2009})}\BibitemShut {NoStop}%
\bibitem [{\citenamefont {Salvador}\ \emph {et~al.}(2014)\citenamefont
  {Salvador}, \citenamefont {Bartumeus}, \citenamefont {Levin},\ and\
  \citenamefont {Ryu}}]{salvador2014mechanistic}%
  \BibitemOpen
  \bibfield  {author} {\bibinfo {author} {\bibfnamefont {L.~C.~M.}\
  \bibnamefont {Salvador}}, \bibinfo {author} {\bibfnamefont {F.}~\bibnamefont
  {Bartumeus}}, \bibinfo {author} {\bibfnamefont {S.~A.}\ \bibnamefont
  {Levin}}, \ and\ \bibinfo {author} {\bibfnamefont {W.~S.}\ \bibnamefont
  {Ryu}},\ }\href@noop {} {\bibfield  {journal} {\bibinfo  {journal} {Journal
  of The Royal Society Interface}\ }\textbf {\bibinfo {volume} {11}},\ \bibinfo
  {pages} {20131092} (\bibinfo {year} {2014})}\BibitemShut {NoStop}%
\bibitem [{\citenamefont {Bir{\'o}}\ and\ \citenamefont
  {Jakov{\'a}c}(2005)}]{biro2005power}%
  \BibitemOpen
  \bibfield  {author} {\bibinfo {author} {\bibfnamefont {T.~S.}\ \bibnamefont
  {Bir{\'o}}}\ and\ \bibinfo {author} {\bibfnamefont {A.}~\bibnamefont
  {Jakov{\'a}c}},\ }\href@noop {} {\bibfield  {journal} {\bibinfo  {journal}
  {Physical review letters}\ }\textbf {\bibinfo {volume} {94}},\ \bibinfo
  {pages} {132302} (\bibinfo {year} {2005})}\BibitemShut {NoStop}%
\bibitem [{\citenamefont {Lubashevsky}\ \emph {et~al.}(2009)\citenamefont
  {Lubashevsky}, \citenamefont {Friedrich},\ and\ \citenamefont
  {Heuer}}]{lubashevsky2009realization}%
  \BibitemOpen
  \bibfield  {author} {\bibinfo {author} {\bibfnamefont {I.}~\bibnamefont
  {Lubashevsky}}, \bibinfo {author} {\bibfnamefont {R.}~\bibnamefont
  {Friedrich}}, \ and\ \bibinfo {author} {\bibfnamefont {A.}~\bibnamefont
  {Heuer}},\ }\href@noop {} {\bibfield  {journal} {\bibinfo  {journal}
  {Physical Review E}\ }\textbf {\bibinfo {volume} {79}},\ \bibinfo {pages}
  {011110} (\bibinfo {year} {2009})}\BibitemShut {NoStop}%
\bibitem [{\citenamefont {Reynolds}\ \emph {et~al.}(2013)\citenamefont
  {Reynolds}, \citenamefont {Schultheiss},\ and\ \citenamefont
  {Cheng}}]{reynolds2013levy}%
  \BibitemOpen
  \bibfield  {author} {\bibinfo {author} {\bibfnamefont {A.~M.}\ \bibnamefont
  {Reynolds}}, \bibinfo {author} {\bibfnamefont {P.}~\bibnamefont
  {Schultheiss}}, \ and\ \bibinfo {author} {\bibfnamefont {K.}~\bibnamefont
  {Cheng}},\ }\href@noop {} {\bibfield  {journal} {\bibinfo  {journal}
  {Behavioral Ecology and Sociobiology}\ }\textbf {\bibinfo {volume} {67}},\
  \bibinfo {pages} {1219} (\bibinfo {year} {2013})}\BibitemShut {NoStop}%
\bibitem [{\citenamefont {Korobkova}\ \emph {et~al.}(2004)\citenamefont
  {Korobkova}, \citenamefont {Emonet}, \citenamefont {Vilar}, \citenamefont
  {Shimizu},\ and\ \citenamefont {Cluzel}}]{korobkova2004molecular}%
  \BibitemOpen
  \bibfield  {author} {\bibinfo {author} {\bibfnamefont {E.}~\bibnamefont
  {Korobkova}}, \bibinfo {author} {\bibfnamefont {T.}~\bibnamefont {Emonet}},
  \bibinfo {author} {\bibfnamefont {J.~M.~G.}\ \bibnamefont {Vilar}}, \bibinfo
  {author} {\bibfnamefont {T.~S.}\ \bibnamefont {Shimizu}}, \ and\ \bibinfo
  {author} {\bibfnamefont {P.}~\bibnamefont {Cluzel}},\ }\href@noop {}
  {\bibfield  {journal} {\bibinfo  {journal} {Nature}\ }\textbf {\bibinfo
  {volume} {428}},\ \bibinfo {pages} {574} (\bibinfo {year}
  {2004})}\BibitemShut {NoStop}%
\bibitem [{\citenamefont {Tu}\ and\ \citenamefont
  {Grinstein}(2005)}]{tu2005white}%
  \BibitemOpen
  \bibfield  {author} {\bibinfo {author} {\bibfnamefont {Y.}~\bibnamefont
  {Tu}}\ and\ \bibinfo {author} {\bibfnamefont {G.}~\bibnamefont {Grinstein}},\
  }\href@noop {} {\bibfield  {journal} {\bibinfo  {journal} {Physical review
  letters}\ }\textbf {\bibinfo {volume} {94}},\ \bibinfo {pages} {208101}
  (\bibinfo {year} {2005})}\BibitemShut {NoStop}%
\bibitem [{\citenamefont {Barab{\'a}si}(2005)}]{barabasi2005origin}%
  \BibitemOpen
  \bibfield  {author} {\bibinfo {author} {\bibfnamefont {A.~L.}\ \bibnamefont
  {Barab{\'a}si}},\ }\href@noop {} {\bibfield  {journal} {\bibinfo  {journal}
  {Nature}\ }\textbf {\bibinfo {volume} {435}},\ \bibinfo {pages} {207}
  (\bibinfo {year} {2005})}\BibitemShut {NoStop}%
\bibitem [{\citenamefont {Matth{\"a}us}\ \emph {et~al.}(2009)\citenamefont
  {Matth{\"a}us}, \citenamefont {Jagodi{\v{c}}},\ and\ \citenamefont
  {Dobnikar}}]{matthaus2009coli}%
  \BibitemOpen
  \bibfield  {author} {\bibinfo {author} {\bibfnamefont {F.}~\bibnamefont
  {Matth{\"a}us}}, \bibinfo {author} {\bibfnamefont {M.}~\bibnamefont
  {Jagodi{\v{c}}}}, \ and\ \bibinfo {author} {\bibfnamefont {J.}~\bibnamefont
  {Dobnikar}},\ }\href@noop {} {\bibfield  {journal} {\bibinfo  {journal}
  {Biophysical journal}\ }\textbf {\bibinfo {volume} {97}},\ \bibinfo {pages}
  {946} (\bibinfo {year} {2009})}\BibitemShut {NoStop}%
\bibitem [{\citenamefont {Matth{\"a}us}\ \emph {et~al.}(2011)\citenamefont
  {Matth{\"a}us}, \citenamefont {Mommer}, \citenamefont {Curk},\ and\
  \citenamefont {Dobnikar}}]{matthaus2011origin}%
  \BibitemOpen
  \bibfield  {author} {\bibinfo {author} {\bibfnamefont {F.}~\bibnamefont
  {Matth{\"a}us}}, \bibinfo {author} {\bibfnamefont {M.~S.}\ \bibnamefont
  {Mommer}}, \bibinfo {author} {\bibfnamefont {T.}~\bibnamefont {Curk}}, \ and\
  \bibinfo {author} {\bibfnamefont {J.}~\bibnamefont {Dobnikar}},\ }\href@noop
  {} {\bibfield  {journal} {\bibinfo  {journal} {PloS one}\ }\textbf {\bibinfo
  {volume} {6}},\ \bibinfo {pages} {e18623} (\bibinfo {year}
  {2011})}\BibitemShut {NoStop}%
\bibitem [{\citenamefont {Reynolds}(2011)}]{reynolds2011origin}%
  \BibitemOpen
  \bibfield  {author} {\bibinfo {author} {\bibfnamefont {A.~M.}\ \bibnamefont
  {Reynolds}},\ }\href@noop {} {\bibfield  {journal} {\bibinfo  {journal}
  {Physica A: Statistical Mechanics and its Applications}\ }\textbf {\bibinfo
  {volume} {390}},\ \bibinfo {pages} {245} (\bibinfo {year}
  {2011})}\BibitemShut {NoStop}%
\bibitem [{\citenamefont {Katul}\ \emph {et~al.}(2005)\citenamefont {Katul},
  \citenamefont {Porporato}, \citenamefont {Nathan}, \citenamefont {Siqueira},
  \citenamefont {Soons}, \citenamefont {Poggi}, \citenamefont {Horn},\ and\
  \citenamefont {Levin}}]{katul2005mechanistic}%
  \BibitemOpen
  \bibfield  {author} {\bibinfo {author} {\bibfnamefont {G.~G.}\ \bibnamefont
  {Katul}}, \bibinfo {author} {\bibfnamefont {A.}~\bibnamefont {Porporato}},
  \bibinfo {author} {\bibfnamefont {R.}~\bibnamefont {Nathan}}, \bibinfo
  {author} {\bibfnamefont {M.}~\bibnamefont {Siqueira}}, \bibinfo {author}
  {\bibfnamefont {M.~B.}\ \bibnamefont {Soons}}, \bibinfo {author}
  {\bibfnamefont {D.}~\bibnamefont {Poggi}}, \bibinfo {author} {\bibfnamefont
  {H.~S.}\ \bibnamefont {Horn}}, \ and\ \bibinfo {author} {\bibfnamefont
  {S.~A.}\ \bibnamefont {Levin}},\ }\href@noop {} {\bibfield  {journal}
  {\bibinfo  {journal} {The American Naturalist}\ }\textbf {\bibinfo {volume}
  {166}},\ \bibinfo {pages} {368} (\bibinfo {year} {2005})}\BibitemShut
  {NoStop}%
\bibitem [{\citenamefont {Shaw}\ \emph {et~al.}(2006)\citenamefont {Shaw},
  \citenamefont {Harwood}, \citenamefont {Wilkinson},\ and\ \citenamefont
  {Elliott}}]{shaw2006assembling}%
  \BibitemOpen
  \bibfield  {author} {\bibinfo {author} {\bibfnamefont {M.~W.}\ \bibnamefont
  {Shaw}}, \bibinfo {author} {\bibfnamefont {T.~D.}\ \bibnamefont {Harwood}},
  \bibinfo {author} {\bibfnamefont {M.~J.}\ \bibnamefont {Wilkinson}}, \ and\
  \bibinfo {author} {\bibfnamefont {L.}~\bibnamefont {Elliott}},\ }\href@noop
  {} {\bibfield  {journal} {\bibinfo  {journal} {Proceedings of the Royal
  Society of London B: Biological Sciences}\ }\textbf {\bibinfo {volume}
  {273}},\ \bibinfo {pages} {1705} (\bibinfo {year} {2006})}\BibitemShut
  {NoStop}%
\bibitem [{\citenamefont {Reynolds}(2013)}]{reynolds2013beating}%
  \BibitemOpen
  \bibfield  {author} {\bibinfo {author} {\bibfnamefont {A.~M.}\ \bibnamefont
  {Reynolds}},\ }\href@noop {} {\bibfield  {journal} {\bibinfo  {journal} {The
  American Naturalist}\ }\textbf {\bibinfo {volume} {181}},\ \bibinfo {pages}
  {555} (\bibinfo {year} {2013})}\BibitemShut {NoStop}%
\bibitem [{\citenamefont {Petrovskii}\ \emph {et~al.}(2011)\citenamefont
  {Petrovskii}, \citenamefont {Mashanova},\ and\ \citenamefont
  {Jansen}}]{petrovskii2011variation}%
  \BibitemOpen
  \bibfield  {author} {\bibinfo {author} {\bibfnamefont {S.}~\bibnamefont
  {Petrovskii}}, \bibinfo {author} {\bibfnamefont {A.}~\bibnamefont
  {Mashanova}}, \ and\ \bibinfo {author} {\bibfnamefont {V.~A.}\ \bibnamefont
  {Jansen}},\ }\href@noop {} {\bibfield  {journal} {\bibinfo  {journal}
  {Proceedings of the National Academy of Sciences}\ }\textbf {\bibinfo
  {volume} {108}},\ \bibinfo {pages} {8704} (\bibinfo {year}
  {2011})}\BibitemShut {NoStop}%
\bibitem [{\citenamefont {Nurzaman}\ \emph {et~al.}(2009)\citenamefont
  {Nurzaman}, \citenamefont {Matsumoto}, \citenamefont {Nakamura},
  \citenamefont {Koizumi},\ and\ \citenamefont
  {Ishiguro}}]{nurzaman2009yuragi}%
  \BibitemOpen
  \bibfield  {author} {\bibinfo {author} {\bibfnamefont {S.~G.}\ \bibnamefont
  {Nurzaman}}, \bibinfo {author} {\bibfnamefont {Y.}~\bibnamefont {Matsumoto}},
  \bibinfo {author} {\bibfnamefont {Y.}~\bibnamefont {Nakamura}}, \bibinfo
  {author} {\bibfnamefont {S.}~\bibnamefont {Koizumi}}, \ and\ \bibinfo
  {author} {\bibfnamefont {H.}~\bibnamefont {Ishiguro}},\ }in\ \href@noop {}
  {\emph {\bibinfo {booktitle} {Robotics and Biomimetics, 2008. ROBIO 2008.
  IEEE International Conference on}}}\ (\bibinfo {organization} {IEEE},\
  \bibinfo {year} {2009})\ pp.\ \bibinfo {pages} {806--811}\BibitemShut
  {NoStop}%
\bibitem [{\citenamefont {Zaburdaev}\ \emph {et~al.}(2015)\citenamefont
  {Zaburdaev}, \citenamefont {Denisov},\ and\ \citenamefont
  {Klafter}}]{zaburdaev2015levy}%
  \BibitemOpen
  \bibfield  {author} {\bibinfo {author} {\bibfnamefont {V.}~\bibnamefont
  {Zaburdaev}}, \bibinfo {author} {\bibfnamefont {S.}~\bibnamefont {Denisov}},
  \ and\ \bibinfo {author} {\bibfnamefont {J.}~\bibnamefont {Klafter}},\
  }\href@noop {} {\bibfield  {journal} {\bibinfo  {journal} {Reviews of Modern
  Physics}\ }\textbf {\bibinfo {volume} {87}},\ \bibinfo {pages} {483}
  (\bibinfo {year} {2015})}\BibitemShut {NoStop}%
\bibitem [{\citenamefont {Wigner}(1958)}]{wigner1958distribution}%
  \BibitemOpen
  \bibfield  {author} {\bibinfo {author} {\bibfnamefont {E.~P.}\ \bibnamefont
  {Wigner}},\ }\href@noop {} {\bibfield  {journal} {\bibinfo  {journal} {Annals
  of Mathematics}\ ,\ \bibinfo {pages} {325}} (\bibinfo {year}
  {1958})}\BibitemShut {NoStop}%
\bibitem [{\citenamefont {Voiculescu}(1991)}]{voiculescu1991limit}%
  \BibitemOpen
  \bibfield  {author} {\bibinfo {author} {\bibfnamefont {D.}~\bibnamefont
  {Voiculescu}},\ }\href@noop {} {\bibfield  {journal} {\bibinfo  {journal}
  {Inventiones mathematicae}\ }\textbf {\bibinfo {volume} {104}},\ \bibinfo
  {pages} {201} (\bibinfo {year} {1991})}\BibitemShut {NoStop}%
\bibitem [{\citenamefont {Schenker}\ and\ \citenamefont
  {Schulz-Baldes}(2005)}]{schenker2005semicircle}%
  \BibitemOpen
  \bibfield  {author} {\bibinfo {author} {\bibfnamefont {J.~H.}\ \bibnamefont
  {Schenker}}\ and\ \bibinfo {author} {\bibfnamefont {H.}~\bibnamefont
  {Schulz-Baldes}},\ }\href@noop {} {\bibfield  {journal} {\bibinfo  {journal}
  {arXiv preprint math-ph/0505003}\ } (\bibinfo {year} {2005})}\BibitemShut
  {NoStop}%
\bibitem [{\citenamefont {Erd{\H{o}}s}(2011)}]{erdHos2011universality}%
  \BibitemOpen
  \bibfield  {author} {\bibinfo {author} {\bibfnamefont {L.}~\bibnamefont
  {Erd{\H{o}}s}},\ }\href@noop {} {\bibfield  {journal} {\bibinfo  {journal}
  {Russian Mathematical Surveys}\ }\textbf {\bibinfo {volume} {66}},\ \bibinfo
  {pages} {507} (\bibinfo {year} {2011})}\BibitemShut {NoStop}%
\bibitem [{\citenamefont {Samorodnitsky}\ and\ \citenamefont
  {Taqqu}(1994)}]{samorodnitsky1994stable}%
  \BibitemOpen
  \bibfield  {author} {\bibinfo {author} {\bibfnamefont {G.}~\bibnamefont
  {Samorodnitsky}}\ and\ \bibinfo {author} {\bibfnamefont {M.~S.}\ \bibnamefont
  {Taqqu}},\ }\href@noop {} {\emph {\bibinfo {title} {Stable non-Gaussian
  random processes: stochastic models with infinite variance}}},\ Vol.~\bibinfo
  {volume} {1}\ (\bibinfo  {publisher} {CRC press},\ \bibinfo {year}
  {1994})\BibitemShut {NoStop}%
\bibitem [{\citenamefont {Nolan}(1999)}]{nolan1999fitting}%
  \BibitemOpen
  \bibfield  {author} {\bibinfo {author} {\bibfnamefont {J.~P.}\ \bibnamefont
  {Nolan}},\ }\href@noop {} {\bibfield  {journal} {\bibinfo  {journal}
  {Applications of Heavy Tailed Distributions in Economics, Engineering and
  Statistics, Washington DC}\ } (\bibinfo {year} {1999})}\BibitemShut {NoStop}%
\bibitem [{\citenamefont {Teuerle}\ and\ \citenamefont
  {Jurlewicz}(2009)}]{teuerle2009random}%
  \BibitemOpen
  \bibfield  {author} {\bibinfo {author} {\bibfnamefont {M.}~\bibnamefont
  {Teuerle}}\ and\ \bibinfo {author} {\bibfnamefont {A.}~\bibnamefont
  {Jurlewicz}},\ }\href@noop {} {\bibfield  {journal} {\bibinfo  {journal}
  {Acta Physica Polonica B}\ }\textbf {\bibinfo {volume} {40}},\ \bibinfo
  {pages} {1333} (\bibinfo {year} {2009})}\BibitemShut {NoStop}%
\bibitem [{\citenamefont {Teuerle}\ \emph {et~al.}(2012)\citenamefont
  {Teuerle}, \citenamefont {{\.Z}ebrowski},\ and\ \citenamefont
  {Magdziarz}}]{teuerle2012multidimensional}%
  \BibitemOpen
  \bibfield  {author} {\bibinfo {author} {\bibfnamefont {M.}~\bibnamefont
  {Teuerle}}, \bibinfo {author} {\bibfnamefont {P.}~\bibnamefont
  {{\.Z}ebrowski}}, \ and\ \bibinfo {author} {\bibfnamefont {M.}~\bibnamefont
  {Magdziarz}},\ }\href@noop {} {\bibfield  {journal} {\bibinfo  {journal}
  {Journal of Physics A: Mathematical and Theoretical}\ }\textbf {\bibinfo
  {volume} {45}},\ \bibinfo {pages} {385002} (\bibinfo {year}
  {2012})}\BibitemShut {NoStop}%
\bibitem [{\citenamefont {Szczepaniec}\ and\ \citenamefont
  {Dybiec}(2014)}]{szczepaniec2014stationary}%
  \BibitemOpen
  \bibfield  {author} {\bibinfo {author} {\bibfnamefont {K.}~\bibnamefont
  {Szczepaniec}}\ and\ \bibinfo {author} {\bibfnamefont {B.}~\bibnamefont
  {Dybiec}},\ }\href@noop {} {\bibfield  {journal} {\bibinfo  {journal}
  {Physical Review E}\ }\textbf {\bibinfo {volume} {90}},\ \bibinfo {pages}
  {032128} (\bibinfo {year} {2014})}\BibitemShut {NoStop}%
\bibitem [{\citenamefont {Szczepaniec}\ and\ \citenamefont
  {Dybiec}(2015)}]{szczepaniec2015escape}%
  \BibitemOpen
  \bibfield  {author} {\bibinfo {author} {\bibfnamefont {K.}~\bibnamefont
  {Szczepaniec}}\ and\ \bibinfo {author} {\bibfnamefont {B.}~\bibnamefont
  {Dybiec}},\ }\href@noop {} {\bibfield  {journal} {\bibinfo  {journal}
  {Journal of Statistical Mechanics: Theory and Experiment}\ }\textbf {\bibinfo
  {volume} {2015}},\ \bibinfo {pages} {P06031} (\bibinfo {year}
  {2015})}\BibitemShut {NoStop}%
\bibitem [{\citenamefont {Dybiec}\ and\ \citenamefont
  {Szczepaniec}(2015)}]{dybiec2015escape}%
  \BibitemOpen
  \bibfield  {author} {\bibinfo {author} {\bibfnamefont {B.}~\bibnamefont
  {Dybiec}}\ and\ \bibinfo {author} {\bibfnamefont {K.}~\bibnamefont
  {Szczepaniec}},\ }\href@noop {} {\bibfield  {journal} {\bibinfo  {journal}
  {The European Physical Journal B}\ }\textbf {\bibinfo {volume} {88}},\
  \bibinfo {pages} {184} (\bibinfo {year} {2015})}\BibitemShut {NoStop}%
\bibitem [{\citenamefont {Wosniack}\ \emph {et~al.}(2017)\citenamefont
  {Wosniack}, \citenamefont {Santos}, \citenamefont {Raposo}, \citenamefont
  {Viswanathan},\ and\ \citenamefont {da~Luz}}]{wosniack2017evolutionary}%
  \BibitemOpen
  \bibfield  {author} {\bibinfo {author} {\bibfnamefont {M.~E.}\ \bibnamefont
  {Wosniack}}, \bibinfo {author} {\bibfnamefont {M.~C.}\ \bibnamefont
  {Santos}}, \bibinfo {author} {\bibfnamefont {E.~P.}\ \bibnamefont {Raposo}},
  \bibinfo {author} {\bibfnamefont {G.~M.}\ \bibnamefont {Viswanathan}}, \ and\
  \bibinfo {author} {\bibfnamefont {M.~G.}\ \bibnamefont {da~Luz}},\
  }\href@noop {} {\bibfield  {journal} {\bibinfo  {journal} {PLOS Computational
  Biology}\ }\textbf {\bibinfo {volume} {13}},\ \bibinfo {pages} {e1005774}
  (\bibinfo {year} {2017})}\BibitemShut {NoStop}%
\bibitem [{\citenamefont {Bychuk}\ and\ \citenamefont
  {O’Shaughnessy}(1994)}]{bychuk1994anomalous}%
  \BibitemOpen
  \bibfield  {author} {\bibinfo {author} {\bibfnamefont {O.~V.}\ \bibnamefont
  {Bychuk}}\ and\ \bibinfo {author} {\bibfnamefont {B.}~\bibnamefont
  {O’Shaughnessy}},\ }\href@noop {} {\bibfield  {journal} {\bibinfo
  {journal} {The Journal of chemical physics}\ }\textbf {\bibinfo {volume}
  {101}},\ \bibinfo {pages} {772} (\bibinfo {year} {1994})}\BibitemShut
  {NoStop}%
\end{thebibliography}%
\end{document}